\documentclass[a4paper,11pt]{article}
\usepackage{amsmath,amssymb}
\usepackage{graphicx}
\usepackage{color}
\usepackage{mathrsfs}
\usepackage{bm}
\usepackage{multirow}
\usepackage{booktabs}
\usepackage{authblk}

\newcommand{\Part}[3]{ \frac{ \partial^{#3} #1 }{ \partial #2^{#3} } }
\newcommand{\V}[1]{\bm{#1} } 
\newcommand{\Tr}[1]{ \mathop{\rm Tr}_{ #1 } }

\newcommand{\Ave}[1]{\left\langle {#1} \right\rangle} 
\newcommand{\dAve}[1]{\left\langle \left\langle {#1} \right\rangle \right\rangle} 
\newcommand{\Wt}[1]{ {\widetilde {#1}} } 
\newcommand{\Wh}[1]{\widehat {#1}} 
 
\newcommand{\sgn}[1]{{\rm sgn}\left({#1} \right)}

\newcommand{\Extr}[1]{ \mathop{\rm Extr}_{ #1 } }

\newcommand{\mR}{\mathbb{R}}

\newcommand{\lb}{\left(}
\newcommand{\rb}{\right)}
\newcommand{\lbb}{\left\{}
\newcommand{\rbb}{\right\}}
\newcommand{\lsb}{ \left[ }
\newcommand{\rsb}{ \right] }

\newcommand{\ldbar}{ \left| \left| }
\newcommand{\rdbar}{ \right| \right| }

\newcommand{\Req}[1]{eq.\ (\ref{eq:#1})}
\newcommand{\BReq}[1]{Eq.\ (\ref{eq:#1})}
\newcommand{\NReq}[1]{(\ref{eq:#1})}
\newcommand{\NReqs}[2]{(\ref{eq:#1},\ref{eq:#2})}
\newcommand{\Reqs}[2]{eqs.\ (\ref{eq:#1},\ref{eq:#2})}

\newcommand{\Rfig}[1]{Fig.\ \ref{fig:#1}}

\newcommand{\Lfig}[1]{\label{fig:#1}}
\newcommand{\Leq}[1]{\label{eq:#1}}
\newcommand{\Rsec}[1]{sec.\ \ref{sec:#1}}
\newcommand{\Lsec}[1]{\label{sec:#1}}
\newcommand{\Rapp}[1]{Appendix\ \ref{sec:#1}}
\newcommand{\be}{\begin{eqnarray}}
\newcommand{\ee}{\end{eqnarray}}
\newcommand{\ba}{\begin{array}}
\newcommand{\ea}{\end{array}}
\newcommand{\no}{\nonumber}

\newcommand{\subbe}{\begin{subequations}}
\newcommand{\subee}{\end{subequations}}
\newcommand{\bs}{\backslash}
\newcommand{\mc}[1]{\mathcal{#1}}
\DeclareMathOperator*{\argmin}{arg\,min}
\DeclareMathOperator*{\argmax}{arg\,max}

\title{Cross validation in LASSO and its acceleration}
\author{Tomoyuki Obuchi and  Yoshiyuki Kabashima}
\affil{
Interdisciplinary Graduate School of Science and Engineering,
\\
Tokyo Institute of Technology, Yokohama, Kanagawa, 226-8502, Japan
}

\begin{document}
\maketitle

\begin{abstract}
We investigate leave-one-out cross validation (CV) as a determinator of the weight of the penalty term in the least absolute shrinkage and selection operator (LASSO). First, on the basis of the message passing algorithm and a perturbative discussion assuming that the number of observations is sufficiently large, we provide simple formulas for approximately assessing two types of CV errors, which enable us to significantly reduce the necessary cost of computation. These formulas also provide a simple connection of the CV errors to the residual sums of squares between the reconstructed and the given measurements. Second, on the basis of this finding, we analytically evaluate the CV errors when the design matrix is given as a simple random matrix in the large size limit by using the replica method. Finally, these results are compared with those of numerical simulations on finite-size systems and are confirmed to be correct. We also apply the simple formulas of the first type of CV error to an actual dataset of the supernovae.
\end{abstract}

\section{Introduction}
Extracting rules from data has been at the heart of modern sciences. 
Johannes Kepler discovered his laws of planetary motion by examining the
data of planetary orbits by trial and error, which later led to classical mechanics. 
Max Planck proposed his law of the black body heat radiation for accurately describing experimental data, which played a key role in the discovery of quantum mechanics. 
As these examples imply, rule extraction has relied mainly on human thoughts. 

The never-ending innovation of measurements and experimental techniques is now resulting in the ongoing creation of a large amount of high-dimensional observation data every day. 
This provides us with situations where rule extraction from data is desired considerably more frequently than ever. Although the entire set of DNA sequences of human beings was identified in 2003, considerable effort must still be made in the days ahead for finding out what rules are written in the dataset. 
Worldwide observation networks of global climate are being consolidated, but analyzing 
the observed data in detail is indispensable for understanding the mechanism of global warming. 
Unfortunately, mechanisms underlying the genome and the global climate are considerably more complicated than those of the planetary orbits and the heat radiation. 
This makes it difficult to discover rules only by human thoughts as has been done thus far.

Sparse modeling may be a promising framework for resolving such difficulty~\cite{Okada:13,Rish:14,Mairal:14,Hastie:15}. 
This generally means methods of statistical modeling or machine learning that describe rules by using a large number of parameters and select a ``sparse'' model in which many of the parameters are set to zero by minimizing sparsity-inducing penalties in conjunction with imposing a good fit to the observed data. Modeling methods of this type are preferable in the sense that one can discover 
a simple and reasonable rule in a semi-automatic manner, with little resort to human thoughts,
from a set of many rules that represent various possible relations. 
The least absolute shrinkage and selection operator (LASSO) is a representative method of 
the sparse modeling~\cite{Tibshirani:96,Efron:04}. In this method, many coefficients of large-dimensional linear regression are pruned by the effect of the $\ell_1$ penalty that is defined by the sum of the absolute values of 
the coefficients. This technique has applications in a wide variety of fields, such as image processing~\cite{Wright:09}, ecology~\cite{Elith:06}, genetics~\cite{Schafer:05}, and astronomy~\cite{Kato:12, Uemura:15}. A similar method is known for the signal recovery problem of compressed sensing~\cite{Donoho:06,Candes:05,Candes:06a,Candes:06b}, which exploits the intrinsic sparsity of objective signals for enhancing the signal processing performance~\cite{Donoho:09-1,Donoho:09-2,Kabashima:09,Ganguli:10,Rangan:10,Krzakala:12,Sakata:13,Nakanishi:15}. 

LASSO is, however, required to solve another problem of determining the strength $\lambda$ 
of the penalty term. Cross validation (CV) is a practically useful strategy for handling this task; its basic concept is to evaluate the prediction error by examining the data under control. Smaller values of the CV error are expected to be better to express the generative model of the data. The minimum, if it exists, of the CV error when changing $\lambda$ is thus considered to obtain an optimal value of $\lambda$. Unfortunately, this reasonable strategy is not well controlled because the behavior of the CV error itself is not fully understood. In particular, there are several variants in the definition of the CV error, each of which can exhibit a different behavior and choose a different optimal value of $\lambda$. Further, conducting CV in a naive manner incurs high computational costs, which makes it difficult to systematically study the behavior of these variants. Even worse, this computational difficulty sometimes forces certain compromises such as scaling down the system size, usage of uncontrolled approximations, or even modifications in research plans.

Given the situation, in this study, we treat leave-one-out (LOO) CV and investigate two types of CV errors, to clarify their properties. Efficient formulas to calculate these two errors are proposed by using belief propagation (BP) in computer science or the cavity method in statistical mechanics~\cite{Yedidia:03,ADVA,Seung:92}, in a perturbative manner. A similar formula has also been proposed for Bayesian learning of simple perceptrons in~\cite{Opper:96}. Our derivation is analogous to that of the approximate message passing (AMP) algorithm~\cite{Donoho:09-2,Rangan:10,Krzakala:12,Kabashima:03}. The resultant formulas have two advantages: The computational cost of the resultant algorithm is considerably reduced from that of the naive algorithm; this reveals a simple connection of the CV errors to the residual sums of squares (RSSs) between the reconstructed and the given measurements, in the large system limit.

In response to this second finding, we analytically assess the two CV errors and the corresponding RSSs to reveal their general properties, in the large system limit under the assumption that the measurement matrix is a random matrix, each component of which is independently identically distributed (i.i.d.) from the zero-mean normal distribution. It is commonly found that both the CV errors exhibit their unique minimums as $\lambda$ changes, but the locations of the minimums are different, and hence, the chosen ``optimal'' values of $\lambda$ are discriminably different between the two CV errors. We compare these two values of $\lambda$ by using the so-called receiver operating characteristic (ROC) curve and compare them to the so-called Younden's index, to find that in the weak noise case, the second CV error chooses a more preferable value of $\lambda$ than the first one. This can be attributed to the fact that the first error tends to overestimate the false positive ratio. Unfortunately, however, our analytical result also clarifies that the above simple formulas derived by the BP in a perturbative manner are not applicable to the second CV error. This is understood by an intricate discussion on the change of the chosen variables in the leave-one-out procedure. These findings are confirmed by numerical experiments on finite-size systems, and our formula is clarified to work well for moderate-size systems.

The rest of this paper is organized as follows: In \Rsec{Problem}, we state LASSO in the context of compressed sensing and explain the LOO CV. In \Rsec{Message}, we explain the application of the cavity method to the evaluation of the CV errors in the LOO CV, clarifying the relation between the CV errors and the RSSs. In \Rsec{Analytic}, we present the analytical result in the case of a random-observation matrix. In \Rsec{Comparison}, we show the result of numerical experiments to support our algorithm and analytical results. An application of the proposed method to the Type Ia supernova data is also presented in this section. The last section is devoted to the conclusion. 

\section{Problem setting}\Lsec{Problem}
Here, we state our problem setting and summarize the quantities of interest. These quantities are analyzed in the subsequent sections to clarify the behavior of CV errors in the LOO CV procedure.

\subsection{Compressed sensing based on LASSO}
In this paper, we introduce LASSO in the context of compressed sensing. 
Let us suppose that a vector $\V{y}\in \mR^{M}$ of measurement is generated from an unknown signal vector $\hat{\V{x}}\in \mR^N$, which is assumed to be sparse, through the following linear process: 
\be
\V{y}=A \hat{\V{x}}+\V{\xi},
\ee
where $A=\lbb A_{\mu i} \rbb_{\mu=1,\cdots,M;\, i=1,\cdots N}  \in \mR^{M\times N}$ represents a measurement (design) matrix and $\V{\xi} \in \mR^M$ denotes the measurement noise each component of which is drawn from the zero-mean normal distribution with variance $\sigma_{\xi}^2$, indicated by $\mathcal{N}(0,\sigma_{\xi}^2)$. The number of measurements $M$ is supposed to be smaller than the dimensions of the representation $N$. Currently, we do not specify the ensemble of $A$ but only assume the scaling of the component as $A_{\mu i}=O(1/\sqrt{N})$. On this condition, we infer the representation $\hat{\V{x}}$ from the given measurement $\V{y}$, by utilizing the sparseness of $\hat{\V{x}}$. The sparsity of $\hat{\V{x}}$ is quantified as follows:
\be
\hat{\rho}=\frac{1}{N}  \sum_{i=1}^{N}|\hat{x}_i|_0 \equiv \frac{1}{N}||\hat{\V{x}} ||_0,
\ee
where $|x|_0$ results in zero if $x=0$ and unity otherwise. The symbol $||\cdot ||_0$ is called $\ell_0$-norm. We can also introduce $\ell_k$-norm as follows: 
\be
|| \V{x} ||_{k}=\lb \sum_{i=1}^{N}|x_i|^{k} \rb^{1/k},
\ee
Given an inferred signal $\V{x}$, we introduce the RSS, $\mathcal{E}$, and the rate, $\epsilon$, as follows:
\be
\mathcal{E}(\V{x})=M\epsilon(\V{x})=\frac{1}{2}||\V{y} - A\V{x} ||^2_2,
\ee

The inference of $\V{x}$ based on LASSO is expressed as follows:
\be
\V{x}^{(1)}(\lambda)=\argmin_{ \V{x} }\lbb \mathcal{E}(\V{x})+\lambda ||\V{x} ||_1\rbb.
\Leq{ell_1}
\ee
Unfortunately, the result is biased; i.e., $\V{x}^{(1)}(\lambda)$ does not agree with $\hat{\V{x}}$ even as $M$ increases because of the presence of the penalty term $\lambda ||\V{x} ||_1$ in \Req{ell_1}. A conventional alternative to $\V{x}^{(1)}$ is obtained by minimizing the RSS on the choice of column vectors associated with $\V{x}^{(1)}$. This can be formulated as follows: 
\be
\V{x}^{(2)}(\lambda)=
\argmin_{ \V{x} }\lbb \frac{1}{2}
\ldbar \V{y} - A \lb \left| \V{x}^{(1)}(\lambda) \right|_0 \circ \V{x} \rb   \rdbar^2_2 \rbb,
\Leq{ell_1+PI}
\ee
where $\circ$ denotes the Hadamard product defined as $ ( \V{v}\circ \V{w} )_i=v_i w_i$, and $\left| \cdot \right|_0$ of a vector is defined as $ ( |\V{v}|_0)_i=|v_i|_0$. Corresponding to the two inferred signals $\V{x}^{(1)}$ and $\V{x}^{(2)}$, we define the two RSSs as follows:
\be
&&
\mathcal{E}_{1}(\lambda)=M\epsilon_1(\lambda)=\mathcal{E}(\V{x}^{(1)}(\lambda)),~~~~~
\mathcal{E}_{2}(\lambda)=M\epsilon_2(\lambda)=\mathcal{E}(\V{x}^{(2)}(\lambda)),
\ee

\subsection{Leave-one-out cross validation}
The idea of LOO CV is as follows: We select one measurement $\mu$ among $M$ measurements and leave it out while inferring the signal, which is formally written as follows:
\be
\V{x}^{(1)\bs \mu}(\lambda)=\argmin_{ \V{x} }
\lbb \frac{1}{2} \sum_{\nu(\neq \mu)}\lb y_{\nu} - \sum_{i=1}^{N}A_{\nu i}x_i \rb ^2+\lambda ||\V{x} ||_1\rbb,
\Leq{ell_1_CV}
\ee
where the symbol $\bs \mu$ denotes the absence of the $\mu$th observation. Using this, we can evaluate a CV error on the $\mu$th measurement as $(1/2)\lb y_{\mu}-\sum_{i=1}^{N}A_{\mu i}x^{(1)\bs \mu}_{i}(\lambda)  \rb^2$. Summing this up for all $\mu=1,\cdots,M$ gives the CV error of the first type:
\be
\epsilon_{\rm LOO}^{(1)}(\lambda)=\sum_{\mu=1}^{M}\frac{1}{2M}\lb y_{\mu}-\sum_{i=1}^{N}A_{\mu i}x^{(1)\bs \mu}_{i}(\lambda)  \rb^2.
\Leq{L_1}
\ee
We can reduce the bias effect by the regularization as \Req{ell_1+PI}, defining
\be
\V{x}^{(2)\bs \mu}(\lambda)=\argmin_{ \V{x} }
\lbb \frac{1}{2} \sum_{\nu(\neq \mu)}\lb y_{\nu} - \sum_{i=1}^{N}A_{\nu i} |x^{(1)\bs \mu}_{i}|_0 x_i \rb ^2 \rbb.
\Leq{ell_1+PI_CV}
\ee
The second type of the CV error can thus be expressed as follows: 
\be
\epsilon_{\rm LOO}^{(2)}(\lambda)=\sum_{\mu=1}^{M}\frac{1}{2M}\lb y_{\mu}-\sum_{i=1}^{N}A_{\mu i}x^{(2)\bs \mu}_{i}(\lambda)  \rb^2.
\Leq{L_2}
\ee
We particularly call these errors the LOO errors (LOOEs); they are central quantities of the analysis described in the subsequent sections.

\subsection{Quantities to examine the quality of inference}
In addition to the LOOEs, we need quantities to examine the quality of inference and compare them with the LOOEs. For this, we introduce the mean squared error (MSE) between the true and the inferred signal. Corresponding to the two inferred signals $\V{x}^{(1)}$ and $\V{x}^{(2)}$ in \Reqs{ell_1}{ell_1+PI}, the two MSEs can be calculated as follows: 
\be
\mathcal{M}_{1}(\lambda)=\frac{1}{N}|| \V{x}^{(1)}- \hat{\V{x}}||^2_2,~~~
\mathcal{M}_{2}(\lambda)=\frac{1}{N}|| \V{x}^{(2)}- \hat{\V{x}}||^2_2.
\Leq{MSE-signal}
\ee
Further, the ratios of the correctly or incorrectly chosen variables are important. The true positive ratio, $TP$, and the false positive ratio, $FP$, are as follows:
\be
TP(\lambda)=
\frac{
\sum_i 
\delta_{1,|\hat{x}_i|_0}
\delta_{1,\left| x^{(1)}_i(\lambda) \right|_0} 
}{
\sum_i \delta_{1,|\hat{x}_i|_0 } 
},
~~~
FP(\lambda)=
\frac{
\sum_i 
\delta_{0,|\hat{x}_i|_0}
\delta_{1, \left|x^{(1)}_i(\lambda) \right|_0} 
}{
\sum_i \delta_{0,|\hat{x}_i|_0 } 
}.
\Leq{TP-FP}
\ee
The so-called ROC curve is a plot of $TP$ against $FP$. This characterizes the quality of inference: If a ROC curve is farther from the straight line $TP=FP$, then the inference is better. Accordingly, we also refer to the so-called Youden's index and define an ``optimal'' value of $\lambda$ chosen according to this index, $\lambda_{\rm YI}$. The definition is as follows: 
\be
&&
D(\lambda)=\min_{x}\lbb (TP(\lambda)-x)^2+(FP(\lambda)-x)^2 \rbb,
\\
&&
\lambda_{\rm YI}=\argmax_{\lambda}D(\lambda).
\ee

\section{Message passing for leave-one-out error}\Lsec{Message}
Suppose that the system size and the number of observations, $N$ and $M$, are sufficiently large, implying that an inferred signal does not change considerably by an addition or deletion of observations. This assumption enables us to perform a perturbative treatment, clarifying the relationship between the LOOEs and the RSSs. 

\subsection{Revisiting approximate message passing}\Lsec{Revisit on}
Let us start by stating a derivation of the known AMP algorithm. This can be done by using the belief propagation (BP) or the cavity method~\cite{Donoho:06,Rangan:10,Krzakala:12,Kabashima:03}. For this, we present a probabilistic formulation for the present problem on the basis of the prescriptions of statistical mechanics. We introduce a Hamiltonian, partition function, and Boltzmann distribution, respectively, as follows:
\be
&&
\mathcal{H}_1(\V{x}) \equiv \mathcal{E}(\V{x})+\lambda ||\V{x}||_1, 
\\
&&
Z_1\lb \beta \rb \equiv \int_{-\infty}^{\infty}  d \V{x}~e^{-\beta \mathcal{H}_1},
\\
&&
P_{1}(\V{x})
=
\frac{e^{-\beta \mathcal{H}_1(\V{x})}}{Z_1}
=
\frac{e^{-\beta \lambda ||\V{x}||_1} \prod_{\mu}\Phi_{\mu}(\V{x})}{Z_1},
\ee
where $\beta$ denotes the inverse temperature and $\Phi_{\mu}$ represents the so-called potential function 
\be
\Phi_{\mu}(\V{x})=e^{-\frac{\beta}{2} \lb y_{\mu}-\sum_i A_{\mu i}x_i  \rb^2 }.
\ee
Note that $\beta$ is independent of the strength of the observation noise $\sigma_{\xi}$, and the limit $\beta \to \infty$ is supposed to be taken after all the calculations as we are interested in the minimum of the above Hamiltonian. We denote the average over the Boltzmann distribution by the angular brackets $\Ave{\cdots }$. BP allows us to calculate the marginal distribution by using two types of messages $(i,j,k=1,\cdots ,N,~\mu,\nu=1,\cdots, M)$ as follows:
\be
&&
\hat{\phi}_{\mu \to i}(x_i)=\int \prod_{j(\neq i)}dx_j~\Phi_{\mu}(\V{x})\prod_{j(\neq i)}\phi_{j\to \mu}(x_j),
\Leq{BP1}
\\
&&
\phi_{i \to \mu}(x_i)=e^{-\beta \lambda |x_i|}\prod_{\nu(\neq \mu)}\hat{\phi}_{\nu \to i}(x_i),
\Leq{BP2}
\ee

A crucial observation to assess \Reqs{BP1}{BP2} is that the exponent of the potential function has a sum of an extensive number of random variables; the central limit theorem thus justifies treating it as a Gaussian variable with the appropriate mean and variance. Hence, according to \Req{BP1}, where $x_i$ is special, we can divide the extensive sum of $\Phi_{\mu}$ as follows: 
\be
\sum_{j=1}^{N}A_{\mu j}x_j
=
A_{\mu i}x_i+\sum_{j(\neq i)}A_{\mu j}x_{j}
\approx
A_{\mu i}x_i+
\sum_{j(\neq i)}A_{\mu j}\bar{x}^{\bs \mu}_{j}+\sqrt{V_{\mu}}z,
\Leq{central}
\ee
where $z$ denotes the zero-mean unit-variance Gaussian variable. The second term on the right-hand side represents the mean of $\sum_{j(\neq i)}A_{\mu j}x_{j}$, and thus, 
\be
\bar{x}^{\bs \mu}_{j}=\Ave{x_{j}}_{\bs \mu},
\ee
where the angular brackets $\Ave{\cdots }_{\bs \mu}$ denote the average over the Boltzmann distribution without the $\mu$th potential function. Now, let us call $\lbb \bar{x}^{\bs \mu}_{j} \rbb$ cavity magnetization. The last term is derived by calculating the variance of $\sum_{j(\neq i)}A_{\mu j}x_{j}$ as follows:
\be
&&
\Ave{
\lb \sum_{j(\neq)i}A_{\mu j}x_{j} \rb^2 
}_{\bs \mu}
-
\Ave{
\sum_{j(\neq)i}A_{\mu j}x_{j} 
}_{\bs \mu}^2
=
\sum_{j,k(\neq i)}A_{\mu j}A_{\mu k}
\lb 
\Ave{x_j x_k}_{\bs \mu}-\Ave{x_j}_{\bs \mu}\Ave{x_k}_{\bs \mu}
\rb 
\no \\ &&
\approx 
\sum_{j,k}A_{\mu j}A_{\mu k}
\lb 
\Ave{x_j x_k}_{\bs \mu}-\Ave{x_j}_{\bs \mu}\Ave{x_k}_{\bs \mu}
\rb 
=
\sum_{j,k}A_{\mu j}A_{\mu k}\frac{\chi^{\bs \mu}_{jk}}{\beta}
\equiv V_{\mu},
\ee
where $\chi^{\bs \mu}_{jk}$ is called the susceptibility matrix (without the $\mu$th observation), which quantifies the correlations between the variables. The terms added at the beginning of the second line have a small contribution of the scaling $O\lb 1/N \rb$; thus, they are negligible, and their addition is justified. 

The application of \Req{central} in \Req{BP1} replaces the integration over $\V{x}$ to that over $z$. Performing this integration yields the following:
\be
\hat{\phi}_{\mu \to i}(x_i)
\propto
e^{
\beta 
\lb 
-\frac{1}{2} \frac{ A_{\mu i}^2 }{ 1+\beta V_{\mu} }x_i^2  
+A_{\mu i} x_i
\frac{
y_{\mu}-\sum_{j(\neq i)} A_{\mu j} \bar{x}^{\bs \mu}_{ j} 
}{
1+\beta V_{\mu} 
} 
\rb 
}.
\ee
Combining this formula with \Reqs{BP1}{BP2}, we can derive a recursion relation of $\lbb \bar{x}^{\bs \mu}_{j} \rbb$, leading to the conventional BP equation.

A more convenient recursion relation can be obtained in terms of the full magnetization $\lbb \bar{x}_{j}=\Ave{x_j} \rbb$ instead of the cavity magnetization $\lbb \bar{x}^{\bs \mu}_{j}=\Ave{x_j}_{\bs \mu} \rbb$. The full marginal distribution of $x_i$ necessarily takes the following modified Gaussian form:
\be
\phi_{i}(x_i)
=e^{-\beta \lambda |x_i|}\prod_{\mu}\hat{\phi}_{\mu \to i}(x_i)
\propto
e^{
\beta 
\lb 
-\frac{1}{2} \Gamma_{i} x_i^2  
+h_i x_i
-\lambda|x_i|
\rb
}.
\ee
where 
\be
&&
\Gamma_i=
\sum_{\mu=1}^{M} \frac{A_{\mu i}^2}{1+\beta V_{\mu} }
,
\\
&&
h_i=
\sum_{\mu=1}^{M} \frac{A_{\mu i}}{1+\beta V_{\mu} }
\lb 
y_{\mu}-\sum_{j(\neq i)}A_{\mu j}\bar{x}^{\bs \mu}_{j} 
\rb
=
\sum_{\mu=1}^{M} A_{\mu i}a_{\mu}+
\sum_{\mu=1}^{M} \frac{ A_{\mu i}^2 }{1+\beta V_{\mu} }\bar{x}^{\bs \mu}_{i}.
\Leq{h_i-1}
\ee
and we set
\be
a_{\mu} \equiv \frac{1}{1+\beta V_{\mu}}\lb y_\mu - \sum_{j}A_{\mu j} \bar{x}^{\bs \mu}_{j} \rb.
\Leq{a_mu-1}
\ee
Hereafter, we call $a_{\mu}$ the cavity residual. We can interpret that the cavity residual $a_{\mu}$ contributes the $\mu$th observation to the effective field $h_i$. Thus, the full magnetization $\bar{x}_{j}$ is obtained from $\bar{x}^{\bs \mu}_{j}$ by adding this contribution of $a_{\mu}$ to the effective field in a perturbative manner. This consideration yields the following:
\be
&&
\bar{x}_{j}
\approx \bar{x}^{\bs \mu}_{j}+\sum_{k}\Part{\bar{x}^{\bs \mu}_{j}}{h_k}{}\Part{h_k}{a_{\mu}}{}a_\mu
=\bar{x}^{\bs \mu}_{j}+\sum_{k}A_{\mu k}\chi^{\bs \mu}_{jk} a_{\mu}.
\Leq{linear}
\ee
Note that we consider only the variation of $h_k$ and do not take into account the change in $\Gamma_k$ when adding the $\mu$th observation. This is because the $\mu$th observation's contribution to $\Gamma_k$ is proportional to $A_{\mu k}^2=O(1/N)$ and is smaller than that to $h_k$ proportional to $A_{\mu k}=O(1/\sqrt{N})$. Basically, our perturbation is connected to the smallness of $A_{\mu k}=O \lb 1/\sqrt{N} \rb$, and only the linear term with respect to $A_{\mu k}$ is important. By substituting \Req{linear} into \Req{a_mu-1} and solving it with respect to $a_{\mu}$, we obtain a simple expression of $a_{\mu}$ in terms of the full magnetization $\lbb \bar{x}_i \rbb_{i}$ 
\be
a_{\mu} \approx y_{\mu}-\sum_{j}A_{\mu j}\bar{x}_{j}. 
\Leq{a_mu-2}
\ee
Similarly, the substitution of \Req{linear} into $m^{\bs \mu}_{i}$ in \Req{h_i-1} yields the following:
\be
h_{i}\approx \sum_{\mu=1}^{M}A_{\mu i}a_{\mu}+\Gamma_{i}\bar{x}_i.
\Leq{h_i-2}
\ee
The neglected terms are proportional to the third and higher orders of $A_{\mu i}$. The last term in \Req{h_i-2} is the well-known Onsager reaction term. 

The full magnetization $\bar{x}_i$ is a function of only $\Gamma_i$ and $h_i$; thus, now, we can calculate $\bar{x}_i$ by recursion if the values of $\lbb \Gamma_i \rbb$ are determined. The functional form of $\bar{x}_i$ becomes simple in the limit $\beta \to \infty$ and is identified with $\V{x}^{(1)}$ in \Req{ell_1}. The fixed point of the AMP is described in this limit as follows:
\subbe
\Leq{AMP^1}
\be
&&
a^{(1)}_{\mu}=y_{\mu} -\sum_{j=1}^{N}A_{\mu j}x^{(1)}_j ,
\\
&&
h^{(1)}_i=\sum_{\mu=1}^{M}A_{\mu i}a^{(1)}_{\mu}+ \Gamma_i x^{(1)}_i,
\\
&&
x^{(1)}_i=\frac{h_i-\lambda~\sgn{h^{(1)}_i} }{\Gamma_i }\Theta\lb |h^{(1)}_i|-\lambda \rb,
\ee
\subee
where $\Theta(x)$ denotes the step function giving $1$ for $x \geq 0$ and $0$ otherwise, and the superscript $^{(1)}$ is attached according to \Req{ell_1}. 

Coefficients $\lbb \Gamma_i \rbb $ can be determined using BP in a similar manner; however, this is not an easy task. Therefore, we omit the derivation and just refer to \cite{Rangan:10,Krzakala:12} for the case of weak correlations where only diagonal terms are important, $\beta V_{\mu}\approx \sum_{j=1}^{N}A_{\mu j}^2 \chi^{\bs \mu}_{jj}$. 

The BP algorithm can also be applied for calculating $\V{x}^{(2)}$. The derivation is essentially the same as \Req{AMP^1}, and the result is as follows:
\subbe
\Leq{AMP^2}
\be
&&
a^{(2)}_{\mu}=y_{\mu} -\sum_{j=1}^{N}A_{\mu j}x^{(2)}_j ,
\\
&&
h^{(2)}_i=\sum_{\mu=1}^{M}A_{\mu i}a^{(2)}_{\mu}+ \Gamma_i x^{(2)}_i,
\\
&&
x^{(2)}_i=\frac{h^{(2)}_i }{\Gamma_i }\Theta\lb |h^{(1)}_i|-\lambda \rb.
\ee
\subee
A crucial difference from \Req{AMP^1} is the dependence on $h^{(1)}_i$. This implies that $\V{x}^{(2)}$ are evaluated by solving \Req{AMP^2} conditioned by the solution of \Req{AMP^1}. As explained later, this difference leads to a difficulty in evaluating $\epsilon_{\rm LOO}^{(2)}$, in contrast to $\epsilon_{\rm LOO}^{(1)}$.

\subsection{Simple formulas of leave-one-out error}\Lsec{Simple formula}
For deriving the AMP, we conducted a perturbation on $\bar{x}^{\bs \mu}_{j}$. In the zero-temperature limit, $\bar{x}^{\bs \mu}_{j}$ is identified with $x^{(1)\bs \mu}_{j}$ in \Req{L_1}. By inserting \Req{linear} in \Req{L_1}, we obtain the following:
\be
\epsilon_{\rm LOO}^{(1)}(\lambda) \approx \frac{1}{2M}
\sum_{\mu=1}^{M}
\lb 
1+\sum_{i,j} A_{\mu i}A_{\mu j}\chi^{\bs \mu}_{ij}
\rb^2
\lb
y_{\mu}-\sum_{i}A_{\mu i}x^{(1)}_{i} 
\rb^2.
\Leq{L_1_app1}
\ee
This has a considerable advantage compared to \Req{L_1}: \BReq{L_1} requires us to solve the optimization problem \NReq{ell_1_CV} $M$ times for evaluating $\epsilon_{\rm LOO}^{(1)}$, but in the case of \Req{L_1_app1}, we need to solve the optimization of \NReq{ell_1} only once. 

The susceptibility matrix $\chi^{\bs \mu}_{ij}$ is the origin of difficulty in the computation of $\lbb \Gamma_i \rbb$. Fortunately, once the solution of \Req{ell_1}, $\V{x}^{(1)}$, is obtained, this can be easily calculated. LASSO separates the variables into two types: Some variables become zero as the solution of \Req{ell_1} and are called {\it inactive}; the other variables take finite values and are {\it active}. Suppose that the active and inactive variables are known and that $\tilde{A}$ is the submatrix corresponding to the active set. The active parts of $\V{x}$ and $\chi^{\bs \mu}$ are also introduced as $\tilde{\V{x}}$ and $\tilde{\chi}^{\bs \mu}$, respectively. To evaluate $\tilde{\chi}^{\bs \mu}$, we need to determine $\tilde{\V{x}}$ in the case without the $\mu$th observation, and denote the solution as $\tilde{\V{x}}^{(1)\bs \mu}$ in accordance with \Req{ell_1_CV}. Correspondingly, we introduce a notation $\V{y}_{\bs \mu}$ expressing $\V{y}$ without the $\mu$th component and $\tilde{A}_{\bs \mu}$ representing $\tilde{A}$ without the $\mu$th row. We assume that the active and inactive sets are stable: A small perturbation $\delta \tilde{\V{h}}$ does not change these sets\footnote{This assumption implies that in the evaluation of the LOOEs, we suppose that the active and inactive sets are unchanged by an addition or deletion of the measurement. Unfortunately, this assumption is not correct; however, the result on $\epsilon_{\rm LOO}^{(1)}$ based on this assumption is correct, while that on $\epsilon_{\rm LOO}^{(2)}$ is incorrect. This difference is shown in \Rsec{LOOEs, MSEs} and explained in \Rsec{Cause of the failure}.}. By using these notations and assumptions, we can now easily obtain $\tilde{\V{x}}$ as follows:
\be
&&
\min_{\tilde{\V{x}} }\lbb \frac{1}{2}||\V{y}_{\bs \mu}-\tilde{A}_{\bs \mu} \tilde{\V{x}} ||_2^2+\lambda ||\tilde{\V{x}}||_1-\delta\tilde{\V{h}}\cdot \tilde{\V{x}} \rbb
\no \\ &&
\Rightarrow
\tilde{\V{x}}^{(1)\bs \mu}=
\lb \tilde{A}^{\rm T}_{\bs \mu}\tilde{A}_{\bs \mu} \rb^{-1}
\lb \tilde{A}^{T}_{\bs \mu} \V{y}_{\bs \mu}+\delta \tilde{\V{h}}
-\lambda \sgn{ \tilde{\V{x}}^{(1)\bs \mu} } 
\Leq{x-tilde1}
\rb.
\ee
Considering the variation with respect to $\delta \tilde{\V{h}}$, we obtain the following: 
\be
\tilde{\chi}^{\bs \mu}=\Part{ \tilde{\V{x}}^{(1)\bs \mu}  }{\V{h}}{}
=
\lb \tilde{A}^{\rm T}_{\bs \mu}\tilde{A}_{\bs \mu} \rb^{-1}.
\Leq{chi}
\ee
All the components of the inactive part of $\chi^{\bs \mu}$ are zero, and thus, the susceptibility matrix is fully calculated. 

\BReq{chi} is seemingly inefficient in that the evaluation of $\tilde{\chi}^{\bs \mu}$ for all $\mu$ requires $M$ inverse operations of a matrix, which is computationally expensive. Fortunately, this computational difficulty can be overcome by using the Sherman--Morrison formula. Denoting $\V{u}^{\rm T}_{\mu}$ as the $\mu$th row vector of $\tilde{A}$, we obtain the following:
\be
\lb \tilde{A}^{\rm T}_{\bs \mu}\tilde{A}_{\bs \mu} \rb^{-1}=
\lb \tilde{A}^{\rm T}\tilde{A} \rb^{-1}
+\frac{ 
\lb\tilde{A}^{\rm T}\tilde{A} \rb^{-1}\V{u}_{\mu} \V{u}^{\rm T}_{\mu} \lb \tilde{A}^{\rm T}\tilde{A} \rb^{-1}
}{
1-\V{u}^{\rm T}_{\mu}\lb\tilde{A}^{\rm T}\tilde{A} \rb^{-1}\V{u}_{\mu}
}.
\Leq{Sherman}
\ee
Hence, $\tilde{\chi}^{\bs \mu}$ is calculated from $\lb \tilde{A}^{\rm T}\tilde{A} \rb^{-1}$ by using a small number of simple products of matrices. The inverse operation appears in $\lb \tilde{A}^{\rm T}\tilde{A} \rb^{-1}$ just once. Eqs. \NReq{L_1_app1}, \NReq{chi}, and \NReq{Sherman} constitute the main result presented in this paper. 

A similar discussion seems to be applicable to the evaluation of \Req{ell_1+PI_CV}. The active and inactive sets are common with $\V{x}^{(1)}$ by definition, and the values of active variables can be calculated as follows: 
\be
\min_{\tilde{\V{x}} }\lbb \frac{1}{2}||\V{y}_{\bs \mu}-\tilde{A}_{\bs \mu} \tilde{\V{x}} ||_2^2-\delta\tilde{\V{h}}\cdot \tilde{\V{x}} \rbb
\Rightarrow
\tilde{\V{x}}^{(2)\bs \mu}=
\lb \tilde{A}^{\rm T}_{\bs \mu}\tilde{A}_{\bs \mu} \rb^{-1}
\lb \tilde{A}^{T}_{\bs \mu} \V{y}_{\bs \mu}+\delta \tilde{\V{h}}\rb.
\Leq{x-tilde2}
\ee
This provides the same susceptibility matrix as \Req{chi}. Hence, the second type of LOOE can also be approximated as follows:  
\be
\epsilon_{\rm LOO}^{(2)}(\lambda) \approx \frac{1}{2M}
\sum_{\mu=1}^{M}
\lb 
1+\sum_{i,j} A_{\mu i}A_{\mu j}\chi^{\bs \mu}_{ij}
\rb^2
\lb
y_{\mu}-\sum_{i}A_{\mu i}x^{(2)}_{i} 
\rb^2.
\Leq{L_2_app1}
\ee
Unfortunately, this approximation is not correct while \Req{L_1_app1} is validated. The detailed reasoning is given in \Rsec{Cause of the failure}.

\subsubsection{In the large-size limit}\Lsec{In the large-size}
In the case of the limit $N \to \infty$, we can obtain an analytic formula for $\tilde{\chi}_{ij}$ under certain conditions and thus, considerably simplify the computations of \Reqs{L_1_app1}{L_2_app1}. We will derive this analytic formula in the next section; here, we will just refer to the result:
\be
\tilde{\chi}^{\bs \mu}_{ij}=\frac{\rho(\lambda)}{\alpha-\rho(\lambda)}\delta_{ij}.
\Leq{chi2}
\ee
where $\rho(\lambda)=(1/N)||\V{x}^{(1)}(\lambda)||_0$ denotes the sparsity of the inferred signal. This result is derived under the assumption that the observation matrix is a random matrix each component of which is i.i.d. from the zero-mean normal distribution $\mathcal{N}(0,N^{-1})$. In such a case, the non-diagonal part of the susceptibility becomes irrelevant as reported in \cite{Rangan:10,Krzakala:12}. The resultant formula of the LOOEs is now very simple
\be
\epsilon_{\rm LOO}^{(k)}(\lambda) \to 
\lb 
\frac{\alpha}{\alpha-\rho(\lambda)}
\rb^2
\frac{1}{2M}
\sum_{\mu=1}^{M}
\lb
y_{\mu}-\sum_{i}A_{\mu i}x^{(k)}_{i} 
\rb^2
=
\lb 
\frac{\alpha}{\alpha-\rho(\lambda)}
\rb^2
\epsilon_{k}(\lambda).
\Leq{L_k_app2}
\ee
Using this formula, in the next section, we examine the behavior of the LOOEs in the limit $N\to \infty$ when $\lambda$ is changed. 

Before moving to the next section, we have two comments to make on \Req{L_k_app2}. One is about the relationship to the Akaike information criterion (AIC). It is known that the LOOE asymptotically agrees with AIC in general. This can be directly seen by expanding \Req{L_k_app2} with respect to $\rho/\alpha$ in the limit $\rho/\alpha \ll 1$: 
\be
2M \epsilon_{\rm LOO}^{(k)}(\lambda)\approx ||\V{y}-A\V{x}^{(k)} ||_2^2+2 \rho \frac{||\V{y}-A\V{x}^{(k)} ||_2^2}{\alpha}.
\ee
The second term is expected to converge to $2N \rho \sigma_{\xi}^2$ in the limit $\rho/\alpha \ll 1$, yielding the expression of AIC. The other comment is about the robustness of \Req{L_k_app2}. We have numerically examined some different ensembles of the observation matrix $A$ and found that the relation \NReq{L_k_app2} between the LOOE and the RSS seems to be fairly robust, while \Req{chi2} is not. We have observed that the non-diagonal components of $\chi$ become important in certain ensembles in which components of $A$ are correlated. These non-diagonal components are complicated, but presumably as a result of nontrivial cancellations, the simple relation $\lb 1+\sum_{ij}A_{\mu i}A_{\mu j}\chi_{ij}^{\bs \mu} \rb^2 \to \lb \alpha/(\alpha-\rho) \rb^2$ seems to hold widely. This finding has an important consequence: Even for realistic situations where the observation matrix is far from the random matrix, \Req{L_k_app2} can be used for accurate approximation of the LOOE. We will revisit this point later when treating the real data of the Type Ia supernovae in \Rsec{Comparison}. Apart from obtaining such a realistic benefit, we should examine whether the relation \NReq{L_k_app2} actually holds for a wider ensemble of $A$ than the simple random matrix ensemble in a more systematic manner, which will be an important future work.

\section{Analytic formulas on the random observation matrix}\Lsec{Analytic}
In this section, in the case of the large-system limit $N \to \infty$, we derive an analytic formula of quantities of interest such as the RSSs $\mathcal{E}_{1,2}$ and MSEs $\mc{M}_{1,2}$, under the assumption that the observation matrix is a random matrix, each component of which is i.i.d. from $\mathcal{N}(0,1/N)$. Considering this limit, we keep the ratio $\alpha=M/N(<1)$ finite along with the sparsity of the true signal $\hat{\rho}=||\hat{\V{x}}||_0/N$.

As noted in \Rsec{Problem}, the noise $\V{\xi}$ is i.i.d. from the normal distribution $\mathcal{N}(0,\sigma_{\xi}^2)$. Moreover, the ensemble of the true signal $\hat{\V{x} }$ is assumed to be the Bernoulli--Gaussian distribution:
\be
P(\V{\hat{x}})=
\prod_{i=1}^{N}
\lbb 
\frac{\hat{\rho}}{\sqrt{2\pi \sigma_x^2}}e^{-\frac{1}{2\sigma_x^2}\hat{x}_i^2} +(1-\hat{\rho})\delta(\hat{x}_i)
\rbb.
\ee
Following statistical mechanical jargon, we call the average over $A$, $\V{\xi}$, and $\hat{\V{x}}$ configurational average, which is represented by square brackets with appropriate subscripts. For example, the average over $\V{\xi}$ and $\hat{\V{x}}$ is written as $\lsb \cdots \rsb_{\V{\xi},\hat{\V{x}}}$.

In this section, we only state the outline of our analysis, give the resultant formulas of the free energies and the related quantities, and show some plots of the quantities of interest. The detailed derivations are deferred to \Rapp{Assessing}. 

\subsection{Outline of analysis}\Lsec{Outline of}
Following the usual prescriptions of statistical mechanics, we define the Hamiltonian $\mathcal{H}$ and the partition function $Z$. According to \Reqs{ell_1}{ell_1+PI}, we define two Hamiltonians as follows:
\be
&&
\mathcal{H}_1(\V{x}^{(1)}|\V{\xi},A,\hat{\V{x}})=\frac{1}{2} \ldbar \bm{y}-A\V{x}^{(1)} \rdbar_2^2
+\lambda \ldbar \V{x}^{(1)} \rdbar_1,\\
\Leq{H_1}
&&
\mathcal{H}_2(\V{x}^{(2)}| \V{x}^{(1)},\V{\xi},A,\hat{\V{x}} )=
\frac{1}{2} \ldbar \bm{y}-A\lb  \left| \V{x}^{(1)} \right|_0 \circ \V{x}^{(2)}  \rb \rdbar_2^2.
\Leq{H_2}
\ee
The corresponding partition functions $Z_1$ and $Z_2$ are defined as follows:
\be
&&
Z_1(\beta| \V{\xi},A,\hat{\V{x}}) 
=\lbb \prod_{i=1}^{N}\int_{-\infty}^{\infty}~dx_i^{(1)}~e^{-\beta \mathcal{H}_1(\V{x}^{(1)}|\V{\xi},A,\hat{\V{x}}  )} \rbb
, 
\Leq{Z_1}
\\ &&
Z_2(\beta|\V{x}^{(1)}, \V{\xi},A,\hat{\V{x}}) 
=
\lbb \prod_{i=1}^{N}\int_{-\infty}^{\infty}~d_{|x_i^{(1)}|_0}x_i~e^{-\beta \mathcal{H}_2(\V{x}| \V{x}^{(1)},\V{\xi},A,\hat{\V{x}}  )} \rbb, 
\Leq{Z_2}
\ee
where 
\be
\int d_{|\xi|_0} x =
\left\{
\begin{array}{cc}
 \int dx & (|\xi|_0=1)  \\   
1  &  (|\xi|_0=0)
\end{array}
\right.,
\ee
and the Boltzmann distributions can be expressed as follows:
\be
p_1(\V{x},\beta| \V{\xi},A,\hat{\V{x}})
=\frac{
e^{-\beta \mathcal{H}_1( \V{x}| \V{\xi},A,\hat{\V{x}} ) }
}{
Z_1(\beta| \V{\xi},A,\hat{\V{x}})
}
\Leq{p_1}
,~~
p_2(\V{x},\beta |\V{x}^{(1)}, \V{\xi},A,\hat{\V{x}})=
\frac{
e^{-\beta \mathcal{H}_2(\V{x}|\V{x}^{(1)}, \V{\xi},A,\hat{\V{x}}) }
}{
Z_2(\beta |\V{x}^{(1)}, \V{\xi},A,\hat{\V{x}})
}.
\Leq{p_2}
\ee
Note that $\V{x}^{(1)}$ conditioning $\mathcal{H}_{2},Z_2$ and $p_2$ is drawn from $p_1$. We assume that the average over these Boltzmann distributions $p_1$ and $p_2$ is denoted by angular brackets with an appropriate subscript. We also introduce double angular brackets denoting the average over both $p_1$ and $p_2$, $\dAve{\cdots}\equiv \Ave{\Ave{\cdots}_2}_1$. The averaged free energies $f_1$ and $f_2$ can thus be defined as follows:
\be
&&
-\beta f_1(\beta) = \frac{1}{N} 
\lsb 
\log Z_1(\beta| \V{\xi},A,\hat{\V{x}}) 
\rsb_{\V{\xi},A,\hat{\V{x}}}
, 
\Leq{f_1-def}
\\ &&
-\beta f_2(\beta) 
= \frac{1}{N} 
\lsb 
\lb \prod_{i=1}^{N}\int_{-\infty}^{\infty}dx_i^{(1)} \rb 
p_1(\V{x}^{(1)},\beta| \V{\xi},A,\hat{\V{x}})
\log Z_2(\beta| \V{x}^{(1)},\V{\xi},A,\hat{\V{x}} ) 
\rsb_{\V{\xi},A,\hat{\V{x}}}
\no \\ &&
=\frac{1}{N} 
\lsb 
\Ave{\log Z_2(\beta| \V{x}^{(1)},\V{\xi},A,\hat{\V{x}} ) }_{1}
\rsb_{\V{\xi},A,\hat{\V{x}}}.
\Leq{f_2-def}
\ee
These are the central objects of our analysis. The rates of RSSs, $\epsilon_1$ and $\epsilon_2$, are derived in the zero-temperature limit. Other quantities of interest can also be derived from the free energies.

To take the configurational average and the average over $\V{x}^{(1)}$ in \Req{f_2-def}, we use the replica method. For the evaluation of $f_2$, we need to introduce two different replica numbers: $n$ for $1/Z_1$ in $p_1$ and $\nu$ for $\log Z_2$. Correspondingly, we introduce the following replica-generating functions: 
\be
&&
\Phi_1(n,\beta)= \lsb Z_1^n \rsb_{ \V{\xi},A,\hat{\V{x}} }, 
\Leq{Phi_1}
\\ &&
\Phi_2(n,\nu,\beta)=
\lsb 
Z_1^{n-1} 
\lb \prod_{i=1}^{N}\int_{-\infty}^{\infty}dx_i^{(1)} \rb 
e^{-\beta \mathcal{H}_1(  \V{x}^{(1)}, \beta |\V{\xi},A,\hat{ \V{x} } )  }
\lb Z_2(\beta| \V{x}^{(1)},\V{\xi},A,\hat{\V{x}}) \rb^\nu  
\rsb_{ \V{\xi},A,\hat{\V{x}} }, 
\Leq{Phi_2}
\ee
We derive the free energies from $\Phi_1$ and $\Phi_2$ by using the following identities:
\be
&&
-\beta f_1=
\lim_{n\to 0}\frac{1}{nN}\log \Phi_1(n,\beta)
=
\lim_{n\to 0}\frac{1}{nN}\log \Phi_2(n,0,\beta)
,
\\ &&
-\beta f_2=
\lim_{n\to 0} \lim_{\nu \to 0} \frac{1}{\nu N}\log \Phi_2(n,\nu,\beta). 
\ee
In the actual procedure, we first assume that $n$ and $\nu$ in \Reqs{Phi_1}{Phi_2} are positive integers, which enables us to calculate the average over the quenched variables as well as the integrations over $\V{x}^{(1)}$. Then, we assume the replica symmetry (RS) in the order parameters explained next, which makes it possible to analytically continue the resultant formulas of $\Phi_2$ with respect to $n$ and $\nu$. Finally, using the analytic continuation, we calculate the limits $n\to 0$ and $\nu\to 0$, yielding the free energies.

\subsection{Free energies, order parameters, and quantities of interest}\Lsec{Free energies,}

\subsubsection{Order parameters and their significance}\Lsec{Order parameters and}
Let us start by summarizing the order parameters characterizing the free energies as follows:
\subbe
\Leq{order parameters}
\be
&&
m_1=\frac{1}{N}\sum_i \Ave{ \hat{x}_i x_i^{(1)} }_1 ,\hspace{0.5cm}
m_2=\frac{1}{N}\sum_i \dAve{\hat{x}_i |x^{(1)}_i|_0x_i^{(2)} },  
\\ &&
Q_1=\frac{1}{N}\sum_i  \Ave{ \lb x^{(1)}_i \rb^2 }_1, \hspace{0.5cm}
Q_2=\frac{1}{N}\sum_i  \dAve{ \lb \left| x^{(1)}_i \right|_0  x^{(2)}_i \rb^2 },
\\ &&
q_1=\frac{1}{N}\sum_i  \Ave{ x^{(1)}_i }_1^2, \hspace{0.5cm}
q_2=\frac{1}{N}\sum_i  \dAve{\left| x^{(1)}_i \right|_0 x^{(2)}_i  }^2,
\\ &&
Q_c=\frac{1}{N}\sum_i \dAve{ x^{(1)}_i x^{(2)}_i }, \hspace{0.5cm}
q_c=\frac{1}{N}\sum_i \dAve{ \Ave{x^{(1)}_i}_1 \left| x^{(1)}_i \right|_0 x^{(2)}_i }.
\ee
\subee
Their meaning is simple: $m_{1,2}$ denote the overlaps with the true signal; $Q_{1,2}$ represent the lengths of the reconstructed signals; $q_{1,2}$ quantify the lengths of the averaged reconstructed signals; and $q_c$ and $Q_c$ indicate the overlaps between the two reconstructed signals, the former reflects the fluctuation, and the latter does not. 
The MSEs \NReq{MSE-signal} are connected to the order parameters as follows:
\be
\mathcal{M}_1=\hat{\rho}\sigma_x^2-2m_1+Q_1,~~
\mathcal{M}_2=\hat{\rho}\sigma_x^2-2m_2+Q_2.
\ee 
For simplicity of notation, we also introduce the following symbols:
\be
\mathcal{M}_c=\hat{\rho}\sigma_x^2-(m_1+m_2)+Q_c,~~
\Wt{\mathcal{M}}_{1,2,c}=\mathcal{M}_{1,2,c}+\sigma_{\xi}^2.
\ee

In the calculation of the zero-temperature limit $\beta \to \infty$, the thermal fluctuation shrinks and the order parameters $Q$ and $q$ have a common value. The following order parameters become finite in the limit $\beta \to \infty$:
\be
\chi_1=\beta (Q_1-q_1),~
\chi_2=\beta (Q_2-q_2),~
\chi_c=\beta (Q_c-q_c).
\ee
From the definition of the order parameters \NReq{order parameters}, we can understand that $\chi_1$ and $\chi_2$ are nothing but the average of the diagonal part of the susceptibility matrix and that an equality $\chi_1=\chi_2$ holds as discussed above.

\subsubsection{$f_1$-related quantities}\Lsec{$f_1$-related quantities}
We are only interested in the zero-temperature limit $\beta \to \infty$, and write the explicit formula of $f_1$ in this limit as follows:
\be
&&
f_1(\beta \to \infty)=
\Extr{\Omega_1}
\Biggl\{
- \frac{1}{2} \hat{Q}_1Q_1+\frac{1}{2} \hat{\chi}_1\chi_1+ \hat{m}_1m_1
\no \\ &&
-\frac{1}{\hat{Q}_1} \lb \hat{\rho}F(\theta_{A} )+ (1-\hat{\rho} ) F(\theta_{I}) \rb
+\frac{\alpha}{2}\frac{\Wt{\mathcal{M}_1}}{1+\chi_1}
\Biggr\}.
\Leq{f_1-zerotemp}
\ee
where $\Omega_1=\lbb \chi_1,Q_1,m_1,\hat{\chi}_1,\hat{Q}_1,\hat{m}_1 \rbb$ and $\Extr{x}$ represents taking an extremization condition with respect to $x$. The variable $\hat{\chi}_1$ is a conjugate order parameter of $\chi_1$, as are the other hatted variables except for $\hat{\rho}$. Further, 
\be
&&
E_k(\theta) \equiv
\int_{\theta}^{\infty} \frac{dz}{\sqrt{2\pi}}~z^k
=
\int_{\theta}^{\infty} Dz~z^k,
\\ &&
F(\theta)=\lambda^2 
\lbb 
E_0(\theta)-\frac{1}{\theta}\frac{ e^{-\frac{1}{2}\theta^2 }  }{\sqrt{2\pi}} +\frac{1}{\theta^2}E_0(\theta)
\rbb,
\\ &&
\theta_{A}=\frac{\lambda}{  \sqrt{ \hat{\chi}_1 +\hat{m}^2_1\sigma_x^2 }  },~
\theta_{I}=\frac{\lambda}{  \sqrt{ \hat{\chi}_1  }  }.
\ee

Next, variational conditions with respect to $\Omega_1$ yield the following equations of state (EOSs) of $\Omega_1$:
\subbe
\Leq{EOS-ell_1}
\be
&&
\hat{\chi}_1=\frac{\alpha \Wt{\mathcal{M}}_1}{(1+\chi_1)^2},
\Leq{EOS-chihat_1}
\\ &&
\hat{Q}_1=\frac{ \alpha }{1+\chi_1},
\Leq{EOS-Qhat_1}
\\ &&
\hat{m}_1= \frac{\alpha }{1+\chi_1}=\hat{Q}_1,
\Leq{EOS-mhat_1}
\\ &&
\chi_1=\frac{2}{ \hat{Q}_1 }\lb \hat{\rho}E_{0}(\theta_{A})+(1-\hat{\rho})E_{0}(\theta_{I} ) \rb  ,
\Leq{EOS-chi_1}
\\ &&
Q_1=\frac{2}{ \hat{Q}_1^2 }\lb \hat{\rho}F(\theta_{A})+(1-\hat{\rho})F(\theta_{I})  \rb ,
\Leq{EOS-Q_1}
\\ &&
m_1=2\frac{ \hat{m}_1 }{ \hat{Q}_1 }\hat{\rho} \sigma_{x}^2E_{0}(\theta_{A}),
\Leq{EOS-m_1}
\ee
\subee
Thus, the sparsity of the reconstructed signal $\rho$, the true positive ratio $TP$, and the false positive ratio $FP$ can be expressed as follows:
\subbe
\be
&&
\rho=2 \lb \hat{\rho}E_{0}(\theta_{A})+(1-\hat{\rho})E_{0}(\theta_{I} ) \rb, \Leq{EOS-rho}
\\ &&
FP=2 E_{0}(\theta_{I}), \Leq{EOS-FP}
\\ &&
TP=2 E_{0}(\theta_{A}). \Leq{EOS-TP}
\ee
\subee

From the EOSs, we get the following simple relations: 
\subbe
\be
&&
\chi_1=\frac{\rho}{\alpha-\rho},
\\ &&
\hat{Q}_1=\alpha-\rho,
\ee 
\Leq{simples1}
\subee
The free energy $f_1$ includes the contribution of the regularization term $\lambda || \V{x}^{(1)} ||_1$. This contribution can be represented by using the relation \NReq{EOS-ell_1} as follows:
\be
\frac{\lambda}{N}\Ave{ || \V{x}^{(1)} ||_1}
=-Q_1\hat{Q}_1+m_1\hat{m}_1+\hat{\chi}_1 \chi_1.
\ee
Subtracting this from $f_1$ and using \Req{EOS-ell_1} again, we obtain the following simple formula of $\epsilon_1$:
\be
\epsilon_1=\frac{1}{\alpha} \lb f_1-\frac{1}{N}\Ave{\lambda || \V{x}^{(1)} ||_1}\rb
=\frac{\hat{\chi}_1}{2\alpha}.
\Leq{epsilon_1-result}
\ee

\subsubsection{$f_2$-related quantities}\Lsec{$f_2$-related quantities}
Similarly, the formula of $f_2$ is as follows:  
\be
&&
\alpha \epsilon_2=f_2(\beta \to \infty)
=
\Extr{\Omega_2} \Biggl\{
\frac{1}{2}\frac{\alpha}{1+\chi_2}
\lb \frac{\chi_c^2}{(1+\chi_1)^2} \Wt{\mathcal{M}}_1-2\frac{\chi_c}{1+\chi_1} \Wt{\mathcal{M}}_c+\Wt{\mathcal{M}}_2\rb
\no \\ &&
-\frac{1}{2}Q_2\hat{Q}_2+\frac{1}{2}\chi_2\hat{\chi}_2+m_2\hat{m}_2+Q_c\hat{Q}_c+\chi_c\hat{\chi}_c
\no \\ &&
-\frac{1}{2\hat{Q}_2}
\Biggl(
\rho\hat{\chi}_2+\hat{m}_2^2m_1+2\hat{Q}_c\lb \hat{\chi}_c \chi_1+\hat{m}_2 m_1\rb +\hat{Q}_c^2 Q_1 
\no \\ &&
+2 \lbb 
\hat{\rho}  ( \hat{\chi}_c+\hat{m}_1\hat{m}_2\sigma_x^2 )^2 G(\theta_{A})
+(1-\hat{\rho})  \hat{\chi}_c^2 G(\theta_{I})
\rbb
\Biggr)
\Biggr\}
\Leq{epsilon_2} 
\ee
where $\Omega_2=\lbb \chi_2,Q_2,m_2,\hat{\chi}_2,\hat{Q}_2,\hat{m}_2,\chi_c,Q_c,\hat{\chi}_c,\hat{Q}_c \rbb$ and 
\be
G(\theta )=
\frac{ \theta^3 }{ \lambda^2 } 
\frac{ e^{-\frac{1}{2}\theta^2 } }{\sqrt{2\pi}},
\ee
The EOSs with respect to $\Omega_2$ are as follows: 
\subbe
\Leq{EOS-ell_1+PI}
\be
&&
\hat{\chi}_2=\frac{\alpha}{(1+\chi_2)^2} 
\lbb \frac{\chi_c^2}{(1+\chi_1)^2} \Wt{\mathcal{M}}_1-2\frac{\chi_c}{1+\chi_1} \Wt{\mathcal{M}}_c+\Wt{\mathcal{M}}_2\rbb,
\Leq{EOS-chihat_2}
\\ &&
\hat{Q}_2=\frac{ \alpha }{1+\chi_2},
\Leq{EOS-Qhat_2}
\\ &&
\hat{m}_2= \frac{\alpha }{1+\chi_2} \lb 1-\frac{\chi_c}{ 1+\chi_1 } \rb,
\Leq{EOS-mhat_2}
\\ &&
\hat{\chi}_c=\frac{\alpha }{(1+\chi_1)(1+\chi_2) }\lb \Wt{\mathcal{M}}_c- \frac{ \chi_c }{ 1+\chi_1 }\Wt{\mathcal{M}}_1\rb ,
\Leq{EOS-chihat_c}
\\ &&
\hat{Q}_c=\frac{\alpha }{1+\chi_2 }\frac{ \chi_c }{ 1+\chi_1 } ,
\Leq{EOS-Qhat_c}
\\ &&
\chi_2=\frac{\rho}{ \hat{Q}_2 } ,
\Leq{EOS-chi_2}
\\ &&
Q_2=\frac{1}{ \hat{Q}^2 }
\Biggl\{ 
\rho \hat{\chi}_2
+\hat{m}_2^2 m_1
+ 2\hat{Q}_c \lb \hat{\chi}_c \chi_1+\hat{m}_2 m_1 \rb +\hat{Q}_c^2 Q_1
\no \\ &&
+2 \lbb 
\hat{\rho}  ( \hat{\chi}_c+\hat{m}_1\hat{m}_2\sigma_x^2 )^2 G(\theta_{A})
+(1-\hat{\rho})  \hat{\chi}_c^2 G(\theta_{I})
\rbb
\Biggr\},
\Leq{EOS-Q_2}
\\ &&
m_2=
\frac{ 1 }{ \hat{Q}_2 }
\lbb 
 m_1(\hat{m}_2+\hat{Q}_c)+2\hat{m}_1\sigma_y^2(\hat{\chi}_c+\hat{m}_1\hat{m}_2\sigma_x^2)G(\theta_{A})
\rbb,
\Leq{EOS-m_2}
\\ &&
\chi_c=
\frac{ 1 }{ \hat{Q}_2 }
\lbb 
\hat{Q}_c\chi_1
+2\lbb
\hat{\rho}(\hat{\chi}_c+\hat{m}_1\hat{m}_2\sigma_x^2)G(\theta_{A})
+
(1-\hat{\rho})\hat{\chi}_cG(\theta_{I}) 
\rbb
\rbb,
\Leq{EOS-chi_c}
\\ &&
Q_c=
\frac{ 1 }{ \hat{Q}_2 }
\lbb 
\hat{\chi}_c\chi_1+\hat{m}_2m_1+\hat{Q}_cQ_1
\rbb.
\Leq{EOS-Q_c}
\ee
\subee
For the order parameters of $\Omega_1$ appearing in \Req{epsilon_2}, we insert the solutions of the extremization condition considered in \Req{EOS-ell_1}. From the EOSs, we obtain the following simple relations:
\subbe
\be
&&
\chi_2=\chi_1=\frac{\rho}{\alpha-\rho},
\\ &&
\hat{Q}_2=\hat{Q}_1=\alpha-\rho,
\\ &&
\epsilon_2=\frac{\hat{\chi}_2}{2\alpha }.
\Leq{epsilon_2-result}
\ee 
\Leq{simples2}
\subee
As expected, the two susceptibilities $\chi_1$ and $\chi_2$ coincide.

\subsection{LOOEs, MSEs, and ROC curve}\Lsec{LOOEs, MSEs}
Since the non-diagonal part of $\chi_{ij}^{\bs \mu}$ can be neglected and the diagonal part is $\chi=\rho/(\alpha-\rho)$, we have $(1+\sum_{i,j}A_{\mu i}A_{\mu j}\chi_{ij}^{\bs \mu})^2=\lb \frac{\alpha}{\alpha-\rho}\rb^2$. From \Reqs{L_k_app2}{EOS-ell_1} and \NReq{epsilon_1-result}, the first type of LOOE becomes 
\be
\epsilon_{\rm LOO}^{(1)}=\frac{1}{2}\Wt{\mc{M}}_1. 
\Leq{L_1_MSE}
\ee
Hence, the LOOE directly connects to the corresponding MSE and calculating its minimum is meaningful. This result is natural and can be derived from a simple consideration. Given $\V{x}^{(1)\bs \mu}$, let us consider the difference between $y_{\mu}$ and the counter part of the reconstructed data $y^{(1)}_{\mu}=\sum_{i}A_{\mu i}x^{(1)\bs \mu}_{i}$. In the present situation, none of the rows of $A$ and none of the components of $\V{\xi}$ are correlated, and hence, $\V{x}^{(1)\bs \mu}$ is also uncorrelated with $\{ A_{\mu i} \}_{i}$ and $\xi_{\mu}$. This implies that the average of $(y_{\mu}-y^{(1)}_{\mu})^2$ over $\xi_{\mu}$ and $\{ A_{\mu i} \}_{i}$ is as follows:
\be
\lsb (y_{\mu}-y^{(1)}_{\mu})^2 \rsb_{\xi_{\mu}, \{ A_{\mu i} \}_i}
=
\frac{1}{N}\sum_{i=1}^{N} \lb \hat{x}_i-x_{i}^{(1)\bs \mu} \rb^2+\sigma_{\xi}^2
\approx 
\Wt{\mc{M}}_1.
\ee
The last relation follows from the smallness of the difference between $\V{x}^{(1)\bs \mu}$ and $\V{x}^{(1)}$. This relation immediately leads to \Req{L_1_MSE}. This correspondence between the MSE and CV error can be proved in a more rigorous way~\cite{Bayati:10,Homrighausen:14}.

Clearly, this discussion is applied to $\epsilon_{\rm LOO}^{(2)}$ since $\V{x}^{(2)\bs \mu}$ is again uncorrelated with $\{ A_{\mu i}\}_i$ and $\xi_{\mu}$, and $\epsilon_{\rm LOO}^{(2)}$ should become
\be
\epsilon_{\rm LOO}^{(2)}=\frac{1}{2}\Wt{\mc{M}}_2.
\Leq{L_2_MSE}
\ee
However, our calculation based on \Req{L_k_app2}, combined with the replica result \NReqs{EOS-ell_1+PI}{epsilon_2-result}, yields the following:
\be
\epsilon_{\rm LOO}^{(2)} = \frac{1}{2}\lbb \frac{\chi_c^2}{(1+\chi_1)^2} \Wt{\mathcal{M}}_1-2\frac{\chi_c}{1+\chi_1} \Wt{\mathcal{M}}_c+\Wt{\mathcal{M}}_2\rbb ~~({\rm incorrect}).
\Leq{L_2_incorrect}
\ee
Only the last term is desired, but the other two terms appear and persist. \BReq{L_k_app2} thus provides an incorrect approximation of $\epsilon_{\rm LOO}^{(2)}$, in contrast to $\epsilon_{\rm LOO}^{(1)}$. 

To obtain quantitative information, we plot the LOOEs in \Rfig{MSE-LOOE}. 
\begin{figure}[htbp]
\begin{center}
\vspace{0mm}
\includegraphics[width=0.48\columnwidth]{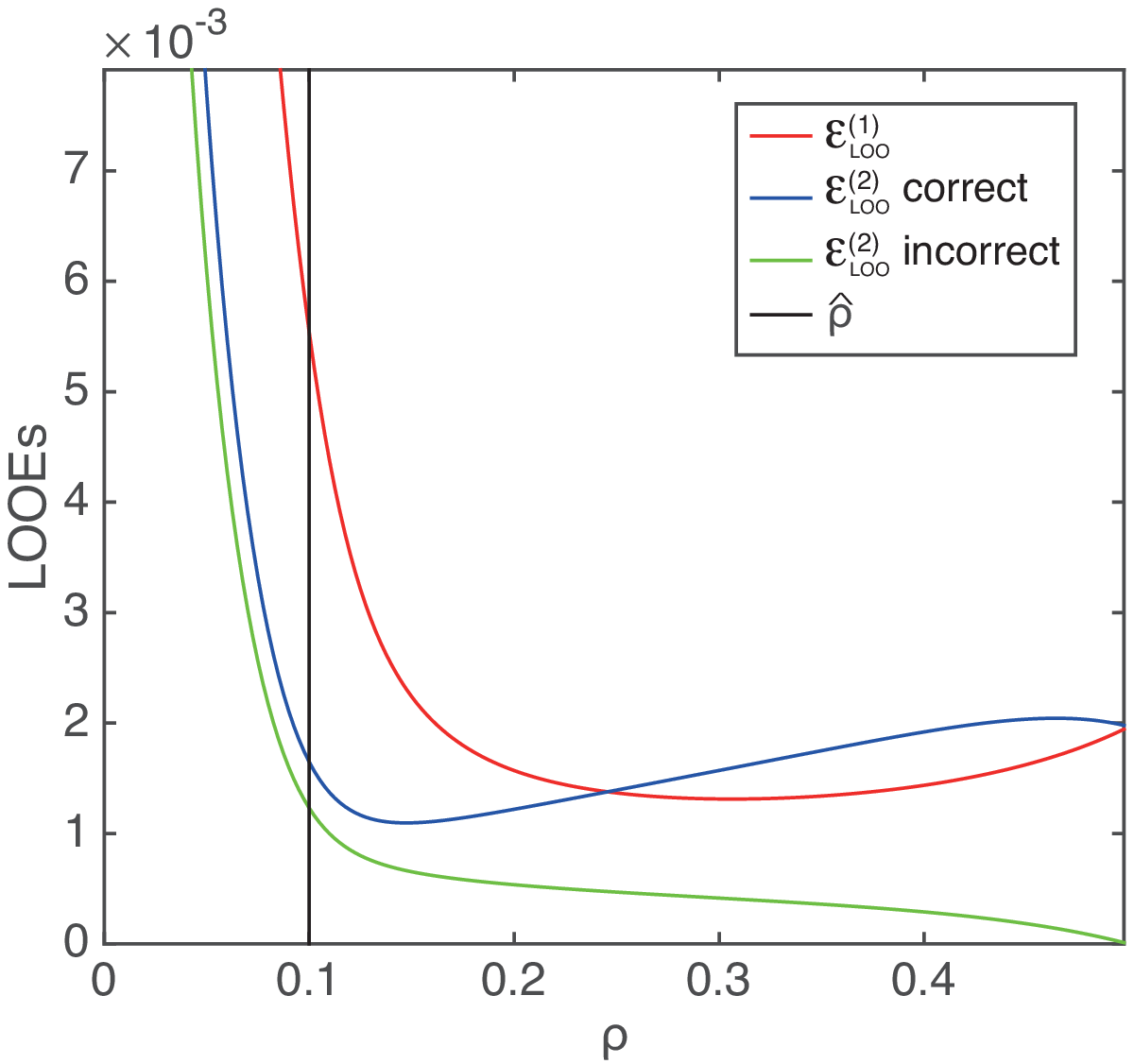}
\includegraphics[width=0.48\columnwidth]{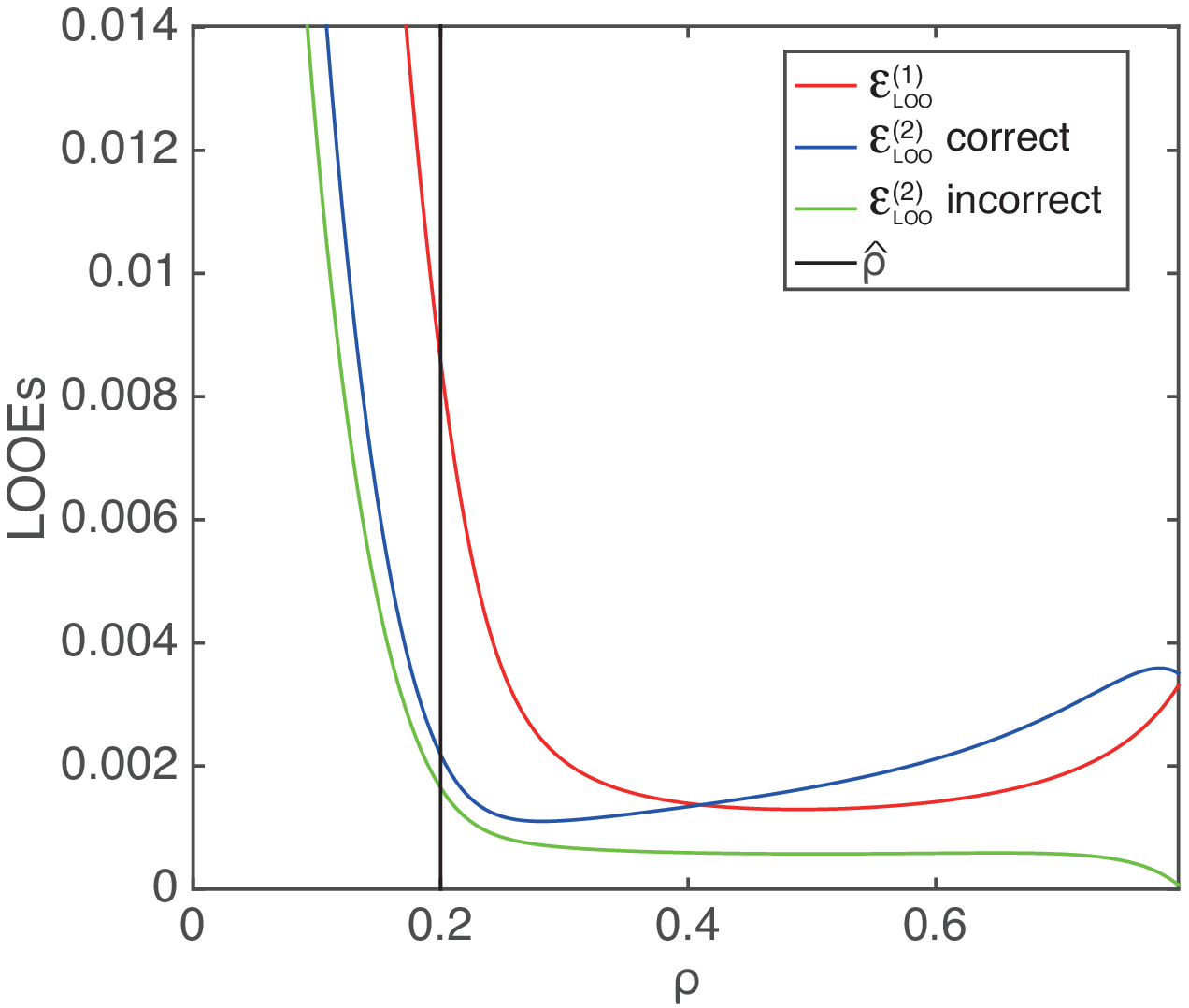}
\vspace{0mm}
\caption{Plots of different estimations of LOOEs: $\epsilon_{\rm LOO}^{(1)}$ (red), the correct estimation of $\epsilon_{\rm LOO}^{(2)}$ (blue), and the incorrect estimation of $\epsilon_{\rm LOO}^{(2)}$ based on \Req{L_2_incorrect}. The parameters are $(\alpha,\hat{\rho})=(0.5,0.1)$ (left) and $(\alpha,\hat{\rho})=(0.8,0.2)$ (right). The signal and noise strengths are commonly set to be $\sigma_{x}^2=1$ and $\sigma_{\xi}^2=0.001$. The difference between the two estimations of $\epsilon_{\rm LOO}^{(2)}$ is not negligible. The minimum value of the correct $\epsilon_{\rm LOO}^{(2)}$ is located at a smaller $\rho$ than that of $\epsilon_{\rm LOO}^{(1)}$ in both the cases. }
\Lfig{MSE-LOOE}
\end{center}
\end{figure}
Three curves are presented in each panel: $\epsilon_{\rm LOO}^{(1)}(=(1/2)\Wt{\mc{M}}_1)$ (red), $\epsilon_{\rm LOO}^{(2)}(=(1/2)\Wt{\mc{M}}_2)$ (blue), and the incorrect evaluation of $\epsilon_{\rm LOO}^{(2)}$ by \Req{L_2_incorrect} (green). The deviation between the blue and the green curves is not negligible and is qualitatively different. The incorrect one converges to zero in the limit $\rho \to \alpha$, but the true one goes to a finite constant identical to the limiting value of $\epsilon_{\rm LOO}^{(1)}$. This implies that the approximation \NReq{L_2_incorrect} is completely useless. Its reasoning will be given later in \Rsec{Cause of the failure}. 

We observe that $\epsilon_{\rm LOO}^{(1)}$ and the correct $\epsilon_{\rm LOO}^{(2)}$ have their unique minimums at certain values of $\rho (< \alpha)$. This implies that the minimums of the LOOEs are good determinators of the value of $\lambda$ since they are connected to the minimums of the MSEs, as shown in \Reqs{L_1_MSE}{L_2_MSE}. To quantify the quality of inference by these two minimums of $\epsilon_{\rm LOO}^{(1)}$ and $\epsilon_{\rm LOO}^{(2)}$, we plot the ROC curves as a plot of $TP$ against $FP$ in \Rfig{ROC}. The $N\to \infty $ solution is denoted by blue points, and the scatter plots of the finite-size simulation with $N=3600$ over $10$ samples are indicated with magenta circles. This simulation is conducted using the built-in ``lasso'' function of MATLAB\textsuperscript{\textregistered}. 
\begin{figure}[htbp]
\begin{center}
\vspace{0mm}
\includegraphics[width=0.48\columnwidth,height=0.4\columnwidth]{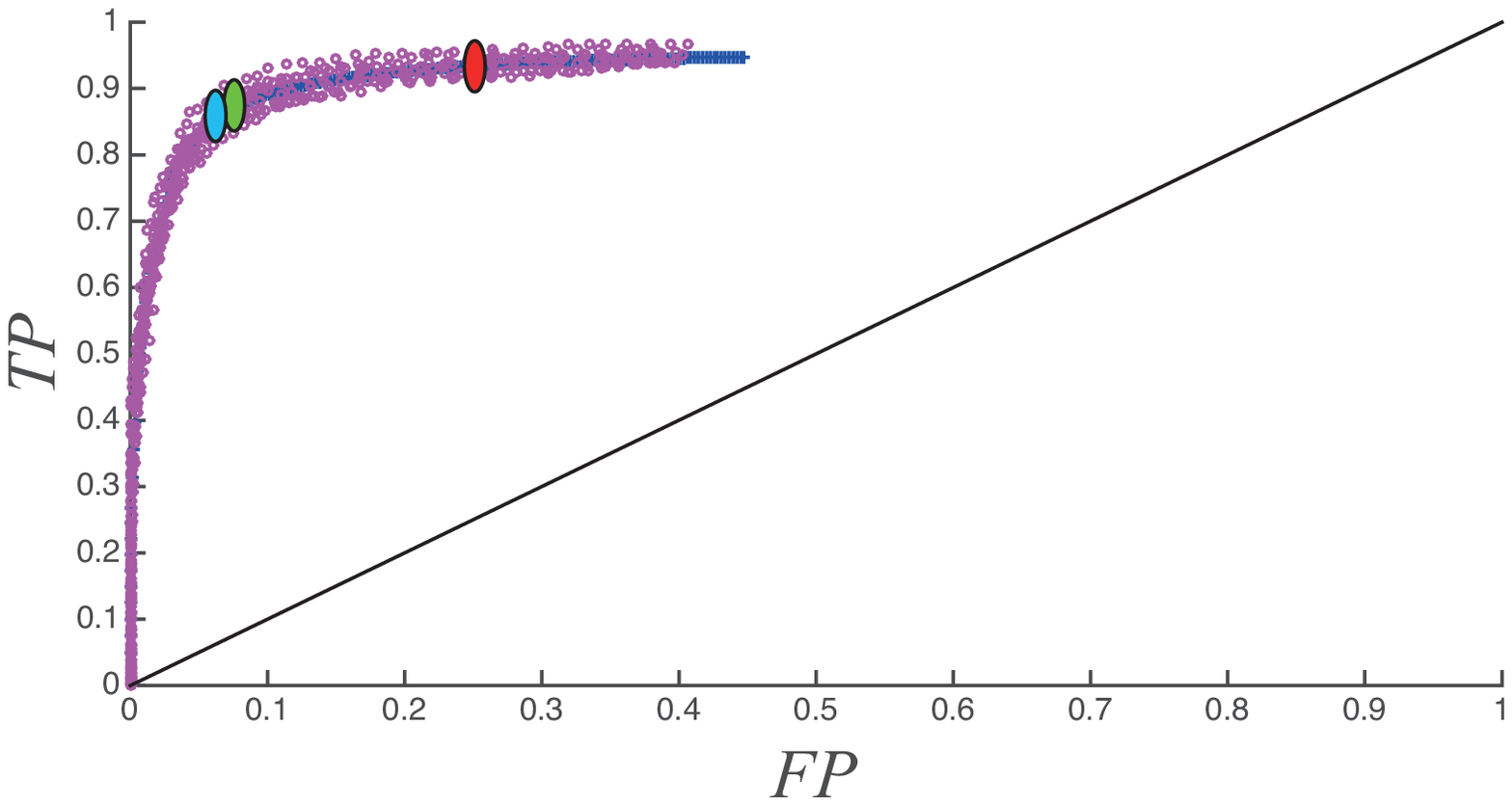}
\includegraphics[width=0.48\columnwidth,height=0.4\columnwidth]{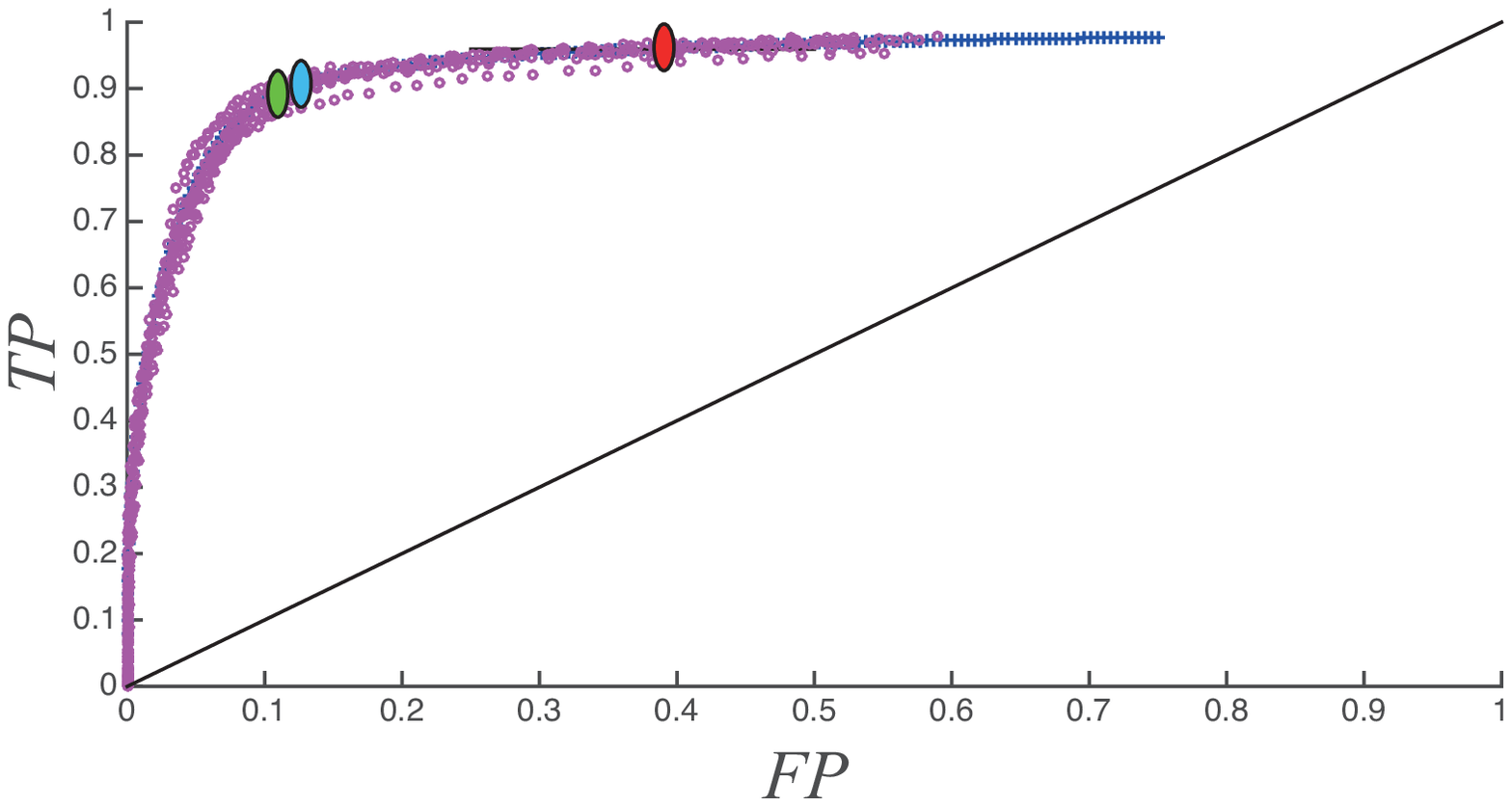}
\vspace{0mm}
\caption{ROC curves for $(\alpha,\hat{\rho})=(0.5,0.1)$ (left) and for $(\alpha,\hat{\rho})=(0.8,0.2)$ (right). The signal and noise strengths are identical to those of \Rfig{MSE-LOOE}. The minimum of $\epsilon_{\rm LOO}^{(1)}$ (red point), Youden's index (green point), and the minimum of $\epsilon_{\rm LOO}^{(2)}$ (blue point) give the coordinate values of $(FP,TP)=(0.24,0.93),~(0.078,0.87)$, and $(0.068,0.86)$ in the left panel, and $(FP,TP)=(0.37,0.96),~(0.11,0.89)$, and $(0.13,0.91)$ in the right panel, respectively. The solution for the limit $N\to \infty$ is denoted by blue points, and the scatter plots of finite-size simulations with $N=3600$ over $10$ samples are indicated with magenta circles.   
}
\Lfig{ROC}
\end{center}
\end{figure}
In the above figure, we mark the points obtained using the minimum of $\epsilon_{\rm LOO}^{(1)}$ as red points, those obtained using Youden's index as green points, and those obtained using the minimum of $\epsilon_{\rm LOO}^{(2)}$ as blue points. The best inference is achieved at the upper-most left point, $(FP,TP)=(0,1)$, and better ROC curves are increasingly skewed to the upper-left direction. The black straight lines denote the $FP=TP$ line and are given as a reference for observing the skewness. \Rfig{ROC} demonstrates that the inference based on LASSO has a good performance. Yet, the points obtained using the minimum values of $\epsilon_{\rm LOO}^{(1)}$, red points, are located a little away from the ``optimal'' values obtained using Youden's index. Meanwhile, the minimums of $\epsilon_{\rm LOO}^{(2)}$ are very close to Youden's optimal values, and in this sense, $\epsilon_{\rm LOO}^{(2)}$ is better than $\epsilon_{\rm LOO}^{(1)}$ for determining $\lambda$. However, note that for obtaining the minimum of $\epsilon_{\rm LOO}^{(2)}$, we have to naively conduct the LOO CV according to its definition since our approximation formulas \NReqs{L_2_app1}{L_k_app2} do not provide reliable estimates of $\epsilon_{\rm LOO}^{(2)}$. Unfortunately, even this naive method faces some difficulties in addition to the computational time. This will be discussed in \Rsec{Application to}. 

Further, we have an additional remark about $\epsilon_{\rm LOO}^{(1)}$. The minimum of $\epsilon_{\rm LOO}^{(1)}$ tends to overestimate the false positive ratio as shown in \Rfig{ROC}. We have checked several different values of the parameters and confirmed that this always holds if the noise is sufficiently weak. An empirical prescription to overcome this is the so-called one-standard error rule, which chooses a larger value of $\lambda$ (corresponding to a smaller $FP$) than that by the minimum of $\epsilon_{\rm LOO}^{(1)}$, by using the error bar of the minimum data point. Hence, it is important to approximate not only the minimum value but also its error bar. Fortunately, our formulas \Reqs{L_1_app1}{L_k_app2} can also provide the error bar of $\epsilon_{\rm LOO}^{(1)}$, which will be demonstrated by a real-data application discussed in \Rsec{Application to}.

\section{Comparison with data on finite size systems}\Lsec{Comparison}
In this section, numerical simulations are presented to examine the validity of our analysis and to determine the finite-size effect. We also apply the proposed method to SuperNova DataBase provided by the Berkeley Supernova Ia program~\cite{Berkeley,Silverman:12} and find that this method reproduces the obtained result~\cite{Uemura:15} considerably faster than the conventional 10-fold CV. 

\subsection{Examination using artificial data}\Lsec{Examination on}
In this subsection, we numerically generate the observation matrix $A$, the true sparse signal $\hat{\V{x}}$, and the noise $\V{\xi}$, matching the assumptions made in our analysis. In all the simulations here, we set $\alpha=0.8, \sigma_x^2=1$, and $\sigma^2_{\xi}=0.001$. Under this condition, once a set of $A,\hat{\V{x}},$ and $\V{\xi}$ is given, we solve \Reqs{ell_1}{ell_1_CV} by using a versatile algorithm of convex optimization, yielding the RSSs and the LOOEs. We generate $N_s=1000$ different samples of the set of $(A,\hat{\V{x}},\V{\xi})$ and adopt the mean value in the samples as our estimate. Error bars are evaluated as standard deviations among the samples divided by $\sqrt{N_s-1}$. In determining active sets after convex optimization, we need to introduce a certain threshold value for each signal component. Here, the threshold value is empirically chosen as $10^{-6}$.

The RSSs for $N=16,32,\cdots, 512$ are given in \Rfig{eps_al08} and compared with the analytical curve of $N\to \infty$.
\begin{figure}[htbp]
\begin{center}
\vspace{0mm}
\includegraphics[width=0.98\columnwidth,height=0.52\columnwidth]{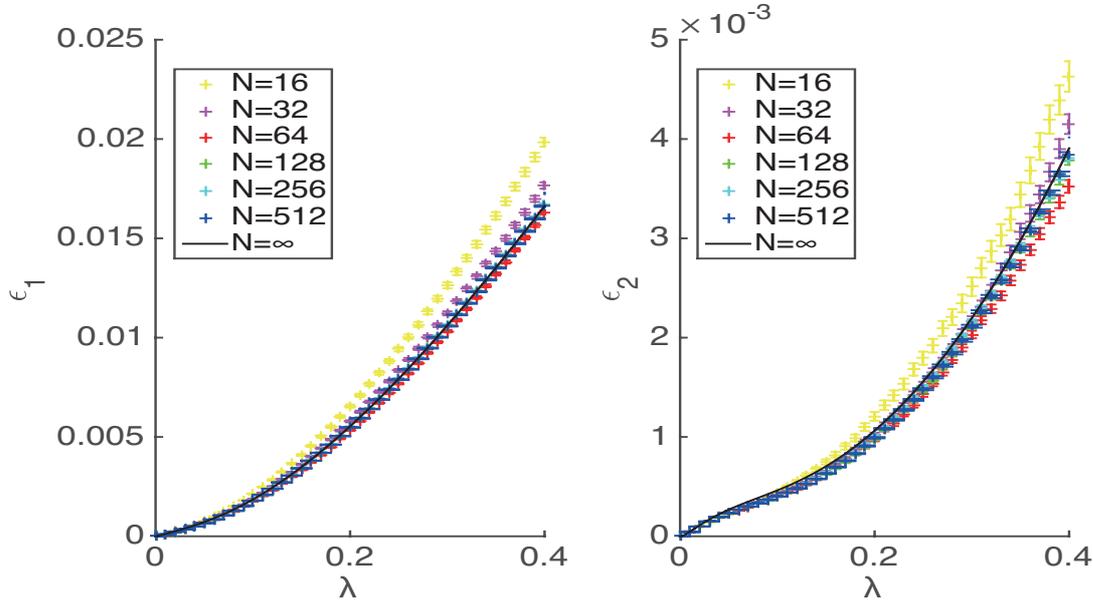}
\vspace{0mm}
\caption{The RSSs $\epsilon_1$ (left) and $\epsilon_2$ (right) are plotted against $\lambda$. The behavior of $\epsilon_2$ is non-monotonic with respect to $N$ in contrast to $\epsilon_1$.  System parameters are set as $\alpha=0.8$, $\hat{\rho}=0.2$, $\sigma_x^2=1$, and $\sigma_{\xi}^2=0.001$. }
\Lfig{eps_al08}
\end{center}
\end{figure}
We find that the finite-size effect is moderate for both $\epsilon_1$ and $\epsilon_2$, but the behavior of $\epsilon_2$ is not monotonic: Up to $N=64$, the numerical values of $\epsilon_2$ tend to become smaller as $N$ increases, but for $N>64$, the values of $\epsilon_2$ start to increase as $N$ increases. Further, the numerical results show a fairly good agreement with the analytical curve, validating our analysis. 

Next, we examine the quality of approximations of $\epsilon_{\rm LOO}^{(1)}$. There are two approximation methods: One is based on \Req{L_1_app1} in conjunction with \Reqs{chi}{Sherman}, which is called {\it Approximation 1} hereafter; the other follows \Req{L_k_app2}, and we call this method {\it Approximation 2}. They are identical in the limit $N\to \infty$, but for finite $N$, there exists a deviation. \Rfig{L1app12_al08} provides $\epsilon_{\rm LOO}^{(1)}$ for $N=16,32,\cdots,512$ evaluated using both Approximations 1 and 2. 
\begin{figure}[htbp]
\begin{center}
\vspace{0mm}
\includegraphics[width=0.98\columnwidth,height=0.52\columnwidth]{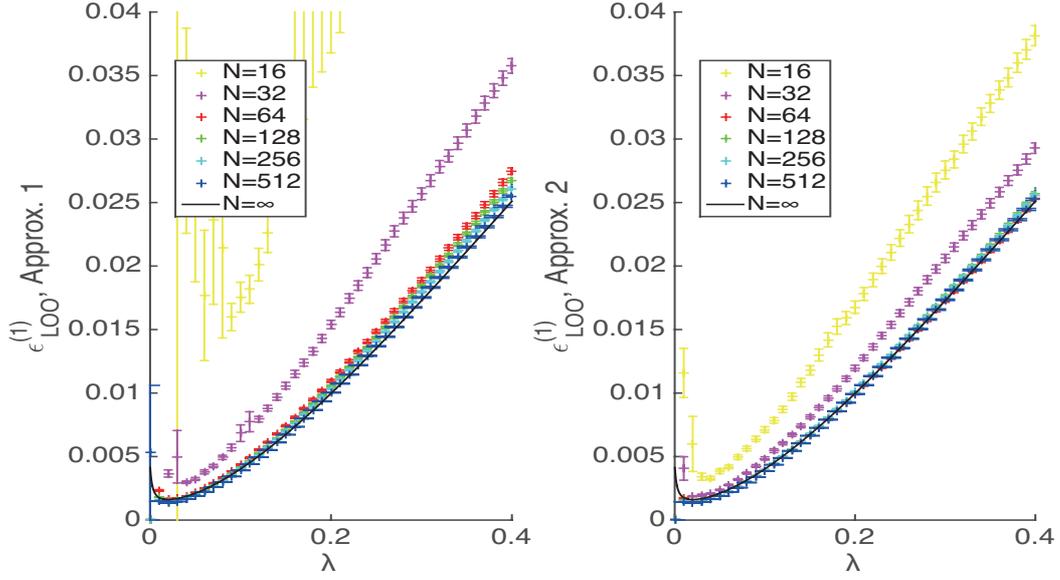}
\vspace{0mm}
\caption{Approximated values of $\epsilon_{\rm LOO}^{(1)}$ based on \Req{L_1_app1} (left) and \Req{L_k_app2} (right) are plotted against $\lambda$. Numerical data (color points) show a fairly good agreement with the analytical black curve for $N\geq 64$. System parameters are identical to those  of \Rfig{eps_al08}.} 
\Lfig{L1app12_al08}
\end{center}
\end{figure}
The finite-size effect is strong and unstable for small $N$, particularly for Approximation 1. This is reasonable because Approximation 1 requires an inversion of the matrix $\tilde{A}^{\rm T}\tilde{A}$. If the number of active variables is close to the number of observations, which is the case for a small $\lambda$, $\tilde{A}^{\rm T}\tilde{A}$ has a mode whose eigenvalue is very close to zero. This leads to the divergence of $(\tilde{A}^{\rm T}\tilde{A})^{-1}$, explaining the drastic change in the LOOEs at small values of $\lambda$ for small sizes. This effect weakens with an increase in the system size, and for $N \geq 64$, the numerical data agrees well with the analytical curve. 

Thirdly, we conduct the LOO CV directly without using any approximation and compare the result with our analytical calculations. The LOO CV is computationally expensive, and we execute it for smaller sizes up to $N=256$. The result is given in \Rfig{CVresult}.
\begin{figure}[htbp]
\begin{center}
\vspace{0mm}
\includegraphics[width=1\columnwidth,height=0.5\columnwidth]{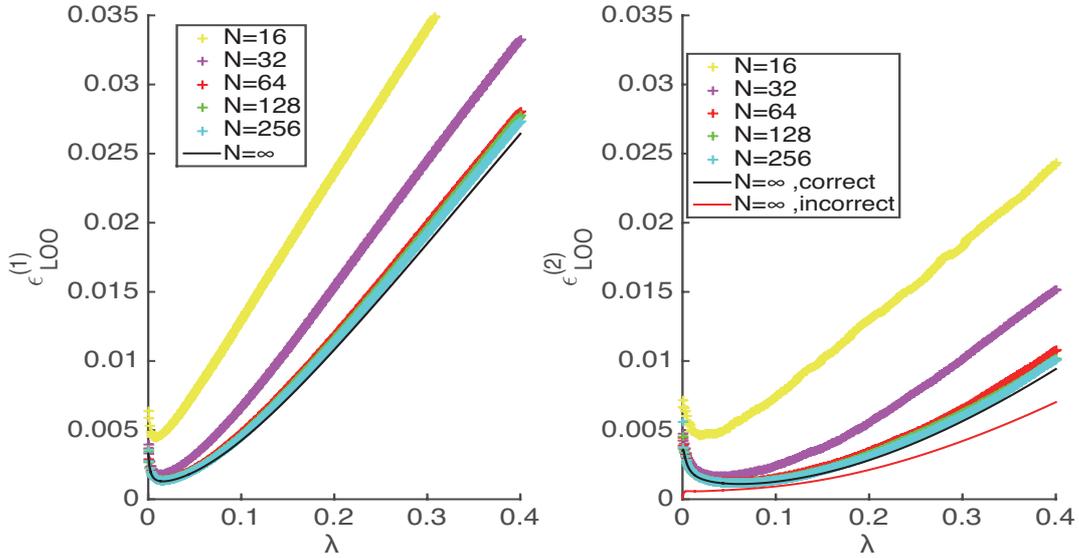}
\vspace{0mm}
\caption{Numerically evaluated LOOEs, $\epsilon_{\rm LOO}^{(1)}$ (left) and $\epsilon_{\rm LOO}^{(2)}$ (right), according to their definitions, are plotted against $\lambda$. For the sake of comparison, analytical curves (black and red solid curves) are drawn. There are two analytical curves for $\epsilon_{\rm LOO}^{(2)}$: The black one is correct and is based on \Req{L_2_MSE}, and the red one is incorrect and is based on \Req{L_2_incorrect}. For $N\geq 64$, the numerical data show a fairly good agreement with the analytical predictions. System parameters are identical to those of \Rfig{eps_al08}.} 
\Lfig{CVresult}
\end{center}
\end{figure}
We can see that our analytical curves (black solid curves) of $\epsilon_{\rm LOO}^{(1)}$ (left) and $\epsilon_{\rm LOO}^{(2)}$ (right) show a fairly good agreement with the direct CV result for $N\geq 64$. In the right panel, we also plot the incorrect prediction based on \Req{L_2_incorrect}, and it exhibits a clear inconsistency with the numerical data. 

Overall, our analytical calculations agree well with the numerical simulations for moderate system sizes, and the approximation formulas \NReqs{L_1_app1}{L_k_app2} provide reliable estimates of $\epsilon_{\rm LOO}^{(1)}$. The benefit of these formulas is their computational ease. For conducting the LOO CV according to its definition, the computational time required is $O(MN_{\rm LASSO})$, where $N_{\rm LASSO}$ denotes the computational time for conducting LASSO once for a desired set of $\lambda$ values, which depend on the system size parameters and the algorithm used. On the other hand, on the basis of our approximation formulas, this is considerably reduced. For example, according to \Req{L_k_app2}, the computational time is $O(N_{\rm LASSO})$. The effect of this acceleration is discussed in \Rsec{Application to}. 

\subsubsection{Cause of the failure on $\epsilon_{\rm LOO}^{(2)}$}\Lsec{Cause of the failure}
Here, we probe the cause of the failure in approximating $\epsilon_{\rm LOO}^{(2)}$ based on \Reqs{L_2_app1}{L_k_app2}, although the same approximation is applicable for $\epsilon_{\rm LOO}^{(1)}$. While deriving \Reqs{L_1_app1}{L_2_app1} and \NReq{L_k_app2}, we assumed that the set of active variables is stable against small perturbations and that the addition or deletion of a row of the observation matrix $A$ just leads to such small perturbations. This assumption is examined here. 

In the absence of the $\mu$th row of $A$, the corresponding fixed-point equations of the AMP become
\subbe
\Leq{AMP^1bsmu}
\be
&&
a^{(1)\bs \mu}_{\nu}=y_{\nu}-\sum_{i}A_{\nu i}x^{(1)\bs \mu}_{i},
\\
&&
h^{(1)\bs \mu}_{i}=\sum_{\nu(\neq \mu)}A_{\nu i}a^{(1)\bs \mu}_{\nu}+\Gamma_i x^{(1)\bs \mu}_{i},
\Leq{h^1bsmu}
\\
&&
x^{(1)\bs \mu}_{i}=\frac{  h^{(1)\bs \mu}_{i}-\lambda \sgn{h^{(1)\bs \mu}_{i}} }{\Gamma_i}\Theta\lb |h^{(1)\bs \mu}_{i}|-\lambda \rb.
\ee
\subee
Let us call this system the $\mu$-cavity system. The difference between \Req{AMP^1bsmu} and \Req{AMP^1} is only the term $A_{\mu i}a_{\mu}=O \lb \sqrt{N}^{-1} \rb$ in \Req{h^1bsmu}. Hence, it is expected that the difference in variables between the full and the $\mu$-cavity systems is small and can be scaled as $O\lb \sqrt{N}^{-1} \rb$. Even if this assumption is correct, it is not trivial to compute the variables of the $\mu$-cavity system. However, the discussion in \Rsec{Simple formula} implies that we estimate this difference as follows:
\be
h^{(1)}_{i}-h^{(1)\bs \mu}_{i} \approx A_{\mu i} a^{(1)}_{\mu}.
\Leq{h-change}
\ee 
Further, we assume that 
\be
\forall{i}, \Theta \lb |h^{(1)}_{i}|-\lambda \rb=\Theta \lb |h^{(1)\bs \mu}_{i}|-\lambda \rb.
\Leq{support-change}
\ee
We have examined these two relations by numerically solving both \Req{AMP^1} and \Req{AMP^1bsmu} independently, and found that the first one is satisfied in a moderate region of $\lambda$ at a certain accuracy, while the second one is incorrect in the entire range of interest of $\lambda$. In fact, it can be proved that the change of active sets in the LOO procedure is inevitable in any sparse algorithm~\cite{Xu:12}. This poses another question: Why is $\epsilon_{\rm LOO}^{(1)}$ well approximated by \Reqs{L_1_app1}{L_k_app2}? 

The violation of \Req{support-change} implies that the active and inactive sets are different for the full and the $\mu$-cavity systems. Some variables belong to the same sets on both the systems and are called ``stable''. The others change the belonging sets and are called ``unstable''. The behavior of the unstable variables is a crucial issue. The effective field $h^{(1)}$ of any unstable variable must satisfy the relation $|h^{(1)}|-\lambda = O\lb \sqrt{N}^{-1} \rb$, since the variation of the effective field, $\Delta h^{(1)\bs \mu} \equiv h^{(1)}- h^{(1)\bs \mu}$, also scales as $O\lb \sqrt{N}^{-1} \rb$ and should be comparable with $|h^{(1)}|-\lambda$. This implies that the coefficient $x^{(1)}$ of any unstable variable is zero or very small as $x^{(1)}\propto |h^{(1)}|-\lambda=O\lb \sqrt{N}^{-1} \rb$. Further, the number of unstable variables is estimated as $N_{\rm uns} \approx |\Delta h^{(1)\bs \mu}| \times N P(h^{(1)})=O\lb \sqrt{N} \rb$, where $P(h^{(1)})$ denotes the distribution of the effective field of the full system and is assumed to be $O(1)$ around $|h^{(1)}|\approx \lambda$. Summarizing these scalings of $N_{\rm uns}$ and the coefficient $x^{(1)}$, we can estimate their contribution as follows:
\be
\sum_{i\in {\rm UNS} }A_{\mu i}x^{(1)}_{i}=O \lb \sqrt{N}\times \sqrt{N}^{-1} \times \sqrt{N}^{-1} \rb=O\lb \sqrt{N}^{-1} \rb \to 0,
\ee
where ${\rm UNS}$ denotes the set of unstable variables. Note that $A_{\mu i}=O\lb \sqrt{N}^{-1} \rb$. The same is true if $x^{(1)}_{i}$ is replaced with $x^{(1)\bs \mu}_{i}$ in the above equation. Hence, the contribution from the unstable variables vanishes in both the full and the $\mu$-cavity systems in the calculation of the cavity residuals $\V{a}^{(1)}$ and $\V{a}^{(1)\bs \mu}$. As a result, our perturbative discussion assuming \Req{support-change} is validated to calculate macroscopic quantities such as $\epsilon_{\rm LOO}^{(1)}$ but cannot correctly compute microscopic information such as ${\rm UNS}$ and the associated coefficients $\{ x_i^{(1)} \}_{i \in {\rm UNS}}$. 

The above reasoning manifests why $\epsilon_{\rm LOO}^{(2)}$ is not correctly evaluated by \Reqs{L_2_app1}{L_k_app2}. Now, the coefficients of unstable variables, $\{ x_i^{(2)} \}_{i\in {\rm UNS}} $, are not proportional to $|h^{(1)}|-\lambda$ and are of $O(1)$, as seen in \Req{AMP^2}. Thus, its contribution is
\be
\sum_{i\in {\rm UNS} }A_{\mu i}x^{(2)}_{i}=O \lb \sqrt{N}\times \sqrt{N}^{-1} \times 1 \rb=O\lb 1 \rb,
\ee
and is not negligible, implying that the solution of the corresponding fixed-point equations is influenced by the unstable variables. Hence, our perturbative discussion does not work even in the calculation of macroscopic quantities.

\subsection{Application to Type Ia Supernova data}\Lsec{Application to}
Here, we apply the proposed method for evaluating $\epsilon_{\rm LOO}^{(1)}$ to the data from SuperNova DataBase provided by the Berkeley Supernova Ia program~\cite{Berkeley,Silverman:12}. Recently, LASSO techniques have been used on a part of the data which is screened by a certain criterion, and a set of important variables known to be significant in explaining the Type Ia supernova data empirically has been reproduced~\cite{Uemura:15}. In this study, the 10-fold CV, which is an alternative to the LOO CV when the number of variables is large and performing the LOO CV is computationally difficult, is used for determining the value of $\lambda$. We calculate $\epsilon_{\rm LOO}^{(1)}$ by using the proposed method on these data and compare the result with that of the 10-fold CV. The system size parameters of the screened data are $M=78$ and $N=276$. 

The left panel of \Rfig{TIadata} shows the plots of $\epsilon_{\rm LOO}^{(1)}$ in Approximations 1 and 2, and the CV error by the 10-fold CV against $\log \lambda$ without the error bars. Clearly, the curves are very similar, and the minimum values of all the curves coincide. 
\begin{figure}[htbp]
\begin{center}
\vspace{0mm}
\includegraphics[width=0.49\columnwidth,height=0.42\columnwidth]{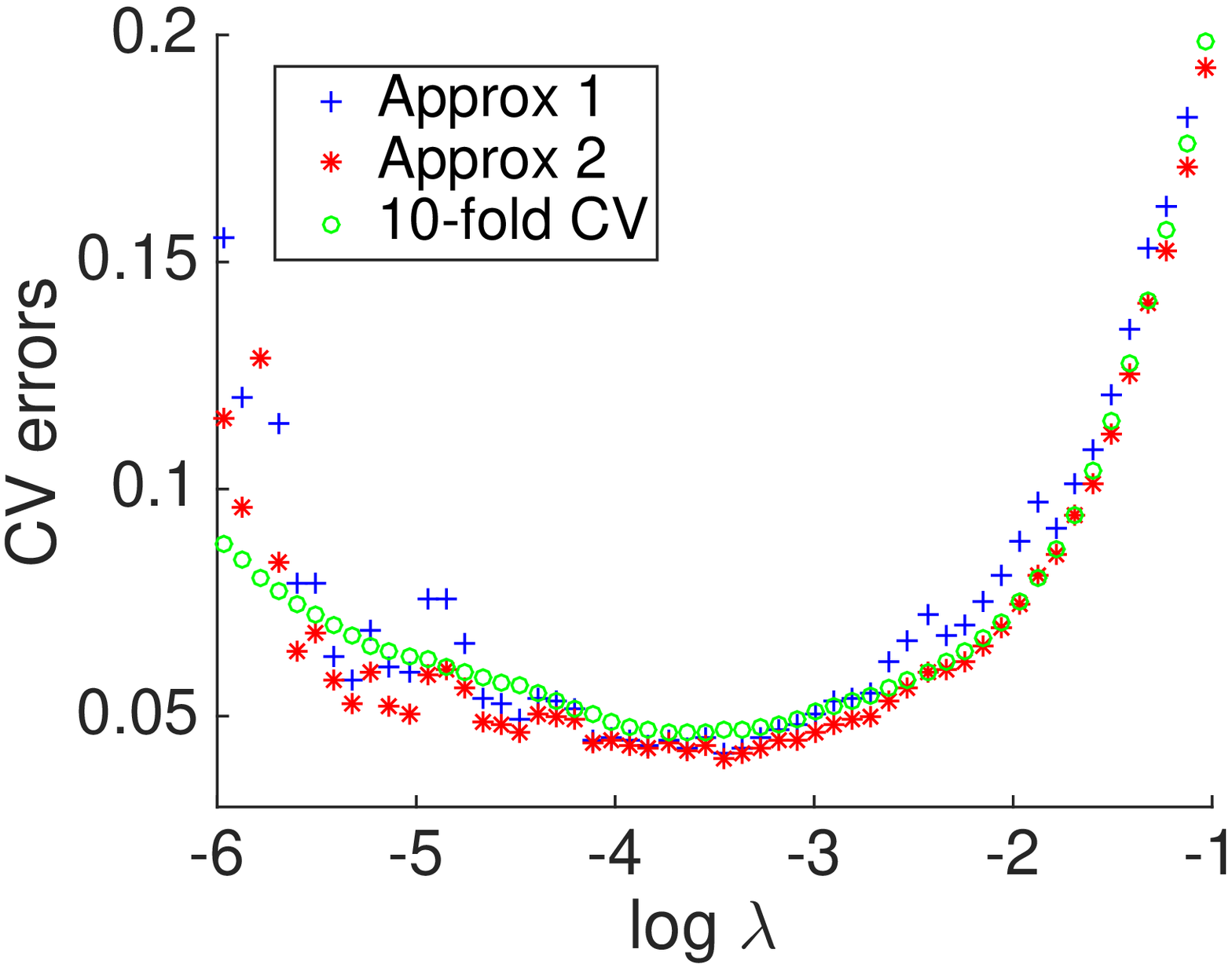}
\includegraphics[width=0.49\columnwidth,height=0.42\columnwidth]{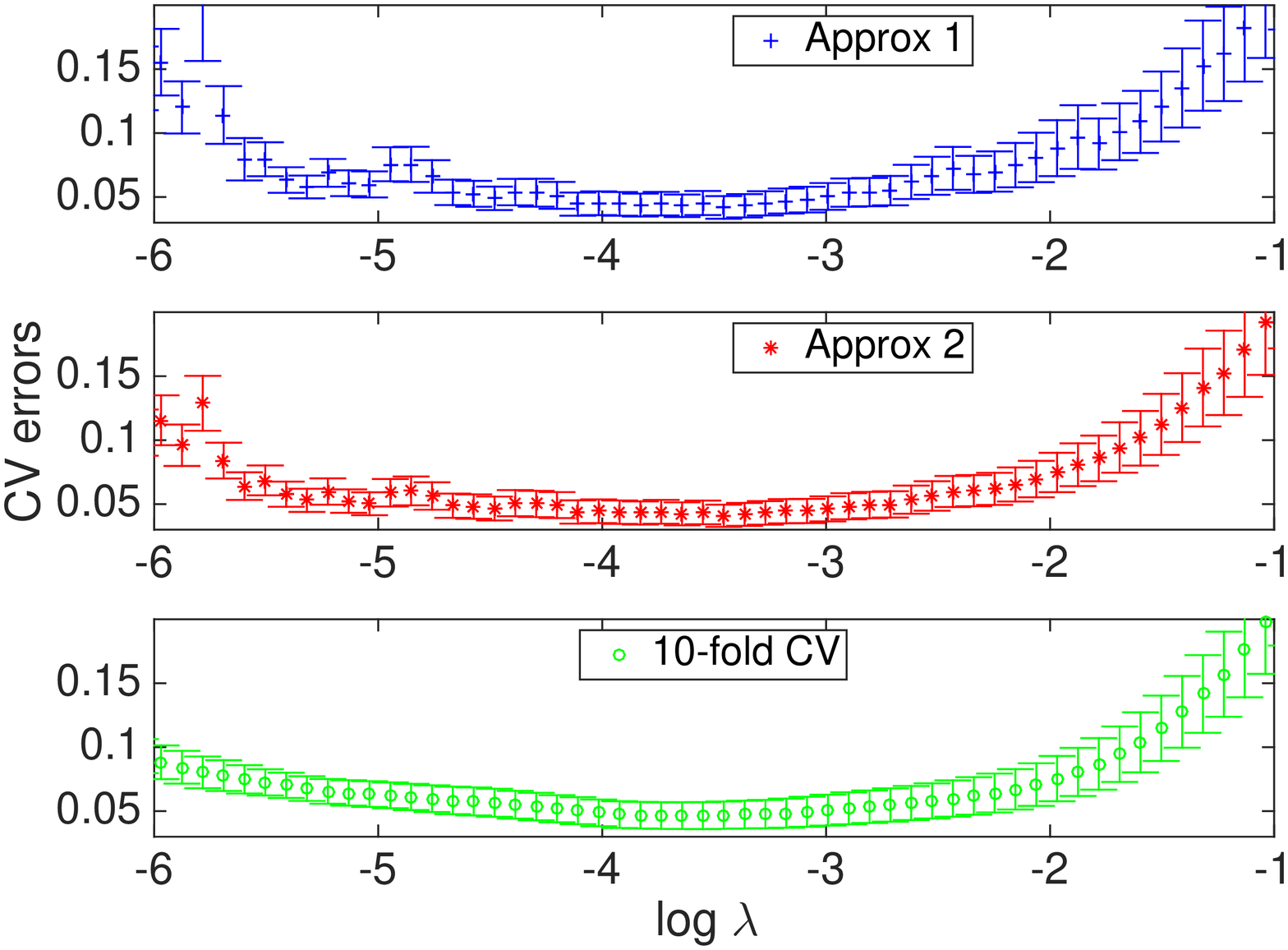}
\vspace{0mm}
\caption{CV errors plotted against $\log \lambda$ for the Type Ia supernova data. Two plots (blue and red) show those obtained using the proposed method, and the other plot (green) illustrates those obtained using the 10-fold CV. The left panel is a direct comparison of the errors without the error bars, and the right panel is the same data with the error bars. All of values are in a good agreement, and the minimums coincide.} 
\Lfig{TIadata}
\end{center}
\end{figure}
We also observe the quantitative similarity of not only the mean values but also the error bars in the right panel of the same figure. The error bars of the 10-fold CV are obtained using a Monte-Carlo resampling, and those of $\epsilon_{\rm LOO}^{(1)}$ evaluated by the proposed method are given by the standard deviation among the $M$ terms of \Req{L_1_app1} or \Req{L_k_app2}, which is justified by a simple resampling argument using the multinomial distribution. Clearly, the largeness of the error bars is quantitatively comparable. Hence, the proposed method reproduces the 10-fold CV result at a very satisfactory level. By applying the one-standard error rule for all the three methods, we obtain ${\rm df}=6$ (${\rm df}$: Number of active variables), which agrees with an empirically validated model, as explained in \cite{Uemura:15}.  

The benefit of the proposed method is apparent in the computational time. The required time for computations in an experiment is 31.6 s for the 10-fold CV, 3.20 s for Approximation 1, and 2.85 s for Approximation 2, the last two of which include the computational time of one run of LASSO. Therefore, the advantage of the proposed method is clear, and the reducing factor is about $10$. If we compare with the LOO CV instead of the 10-fold CV, the factor will be considerably larger. Hence, our approximation is applicable to realistic problems and is very efficient for data with a large dimensionality. Therefore, readers are strongly encouraged to use the presented method.

The observation matrix in the Type Ia supernova data is significantly different from a simple random matrix. Hence, the success of Approximation 2 in the case of these data conversely suggests that the relation $(1+\sum_{i,j}A_{\mu i}A_{\mu j}\chi_{ij}^{\bs \mu} )^2 \approx ( \alpha/(\alpha-\rho) )^2$, which is used for deriving \Req{L_k_app2}, holds rather universally, as discussed at the end of \Rsec{In the large-size}. As mentioned above, Approximation 2 has a relatively low computational time, and its result is stable compared to that of Approximation 1. These facts positively motivate us to find further theoretical evidence for the universality of the relation $(1+\sum_{i,j}A_{\mu i}A_{\mu j}\chi_{ij}^{\bs \mu} )^2 \approx ( \alpha/(\alpha-\rho) )^2$.

Now, finally, we report the behavior of $\epsilon_{\rm LOO}^{(2)}$ on the Type Ia supernova data. The LOO CV according to its definition is conducted, and the result is shown in \Rfig{L2_TIa}.
\begin{figure}[htbp]
\begin{center}
\vspace{0mm}
\includegraphics[width=0.45\columnwidth]{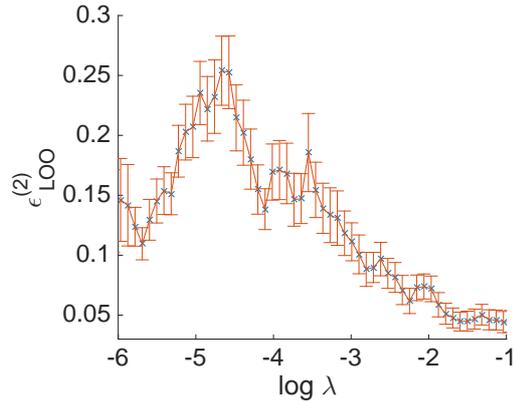}
\vspace{0mm}
\caption{Second type of LOOE $\epsilon_{\rm LOO}^{(2)}$ plotted against $\log \lambda$ for the Type Ia supernova data. } 
\Lfig{L2_TIa}
\end{center}
\end{figure}
The minimum value of $\epsilon_{\rm LOO}^{(2)}$ is located at $\log \lambda \approx -1$, leading to ${\rm df}=1$. This serious reduction of ${\rm df}$ from the case of $\epsilon_{\rm LOO}^{(1)}$ might have a certain meaning. In fact, in \cite{Uemura:15}, an appropriate preprocessing of the same data shows the same serious reduction of ${\rm df}$, which can be attributed to some inconvenient properties of the data such as collinearity and bad statistics. The possibility that $\epsilon_{\rm LOO}^{(2)}$ can diminish the influence of these bad properties is suggested. 

Unfortunately, as seen in the rugged uncontrolled behavior of $\epsilon_{\rm LOO}^{(2)}$, the quality of the presented data is not sufficiently good to judge whether this possibility is plausible or not, although it is natural that ${\rm df}$ is reduced when using $\epsilon_{\rm LOO}^{(2)}$ instead of $\epsilon_{\rm LOO}^{(1)}$, as demonstrated in \Rfig{ROC}. Basically, $\epsilon_{\rm LOO}^{(2)}$ requires considerably better statistics to exhibit a meaningful behavior than $\epsilon_{\rm LOO}^{(1)}$. The reason for this is as follows: When changing $\lambda$, each term in $\epsilon_{\rm LOO}^{(2)}$ consists of several different piecewise constants, and they are not necessarily monotonic. This is because the inferred signal $\V{x}^{(2)\bs \mu}(\lambda)$ shows a sudden change only at certain several discrete points of $\lambda$, where the set of active variables changes, and remains unchanged in-between the neighboring pairs of these discrete points. This is in contrast to $\V{x}^{(1)\bs \mu}(\lambda)$, which changes continuously~\cite{Efron:04,The Elements}. These discrete points change among different terms in $\epsilon_{\rm LOO}^{(2)}$. Hence, $\epsilon_{\rm LOO}^{(2)}$ consists of the sum of the piecewise constants with different jumping points and heights, leading to large error bars and uncontrolled behaviors of $\epsilon_{\rm LOO}^{(2)}$. 

Once this uncontrolled behavior of $\epsilon_{\rm LOO}^{(2)}$ is tamed, we expect an optimal value of $\lambda$ to be chosen by using $\epsilon_{\rm LOO}^{(2)}$ without employing the ad hoc one-standard error rule~\cite{The Elements,Classification,Tibshirani:01}. New ideas are desired for taming. An idea based on bootstrapping can be a good candidate: Increasing the statistics of the present data can diminish the abovementioned discrete behavior. Problems of rather large sizes may not pose this peculiarity in the first place, since the statistics of the LOOE automatically improve with an increase in $M$. In less large but moderate-size problems, the $k$-fold CV for $\epsilon_{\rm LOO}^{(2)}$ with a moderate value of $k$ is worth trying. However, further consideration of $\epsilon_{\rm LOO}^{(2)}$ is desired.

\section{Conclusion}\Lsec{Conclusion}
In this study, we examined the LOO CV as the determinator of the coefficient $\lambda$ of the penalty term in LASSO. We investigated two types of CV errors by using the LOO CV, namely the LOOEs $\epsilon_{\rm LOO}^{(1)}$ and $\epsilon_{\rm LOO}^{(2)}$ corresponding to two different estimators, and for both the errors, we derived simple formulas that significantly reduce the computational cost in their evaluation. This result was derived by using the BP or cavity method and by a perturbative argument assuming that the number of observations was sufficiently large. 

On the basis of this finding, we analytically evaluated the LOOEs by using the replica method when the observation matrix is a simple random matrix. This provided quantitative information about the LOOEs. Further, the locations of the minimums of the two LOOEs were found to be different, and thus, the chosen ``optimal'' values of $\lambda$ are different in general. Both the optimal values were examined by using ROC curves, and the one obtained using the second LOOE $\epsilon_{\rm LOO}^{(2)}$ was found to be preferable. However, a replica analysis clarified that our simple formulas are not useful for accurately approximating $\epsilon_{\rm LOO}^{(2)}$. We need further consideration to design an efficient algorithm for computing $\epsilon_{\rm LOO}^{(2)}$.

The above analytical calculations were compared with numerical simulations on finite-size systems. For small system sizes, there exists a deviation, but for moderate and large system sizes, the agreement between the numerical data and the analytical result is fairly good, and thus, our formulas are validated. We also applied these formulas to real Type Ia supernova data, to find that the proposed method reproduces the known result at a very satisfactory level. The benefit of our formulas is their low computational cost, and the actual reducing factor in the computational time was about 10. Further, the computation of $\epsilon_{\rm LOO}^{(2)}$ according to its definition was conducted on this data set, but it turned out to be difficult to obtain any meaningful result. Larger amounts of data are desired to treat $\epsilon_{\rm LOO}^{(2)}$. 

The proposed method requires the computation of a pre-factor $\sum_{ij}A_{\mu i}A_{\mu j}\chi_{ij}^{\bs \mu}$ and it is in fact a non-trivial task. We had recourse to an analytical formula for this quantity, which can be directly validated in the case where the observation matrix is a simple random matrix but cannot be justified in general. Some of our numerical (unreported) observations support that the analytical formula holds for a wider ensemble of the observation matrix, but deeper theoretical evidence is strongly desired. 

\section*{Acknowledgments}
We would like to express our sincere gratitude to Makoto Uemura and Shiro Ikeda for their helpful discussion on the Type Ia supernova data. The UC Berkeley SNDB is acknowledged for permission of using the data set of \Rsec{Application to}. 
This work was supported by JSPS KAKENHI Grant Numbers 26870185 (TO) and 25120013 (YK).

\appendix
\section{Assessing the generating function $\Phi_2$}\Lsec{Assessing}
For positive integers $n$ and $\nu$, the generating function $\Phi_2(n,\nu)$ can be expressed as follows:
\be
\Phi_2(n,\nu)=
 \Wt{\Tr{\lbb \V{r}_a \rbb_{a=1}^{n} }}  \Tr{\lbb \V{x}_{\alpha} | \V{r}_1 \rbb_{\alpha=1}^{\nu} }
\lsb  
 e^{
 -\frac{1}{2}\beta  
 \lbb 
 \sum_{a=1}^{n}  (d_{\mu a} +\xi_{\mu})^2 + \sum_{\alpha=1}^{\nu}   (\tilde{d}_{\mu \alpha}+\xi_{\mu})^2
\rbb
}
\rsb_{\V{\xi},A,\hat{\V{x}}},
\ee
where 
\be
&&
 \Wt{ \Tr{\V{r}} }=\prod_{i=1}^{N}
 \lbb 
 \int dr_i~e^{-\beta \lambda |r_i|}
 \rbb
 ,\hspace{0.5cm}    
\Tr{\V{x}| \V{r} }=\prod_{i=1}^{N}  
\lbb 
\int d_{|r_i|_0}x_i  \frac{1}{ \sqrt{2\pi} }e^{-\frac{1}{2} x_i^2 }
\rbb.
\\ &&
d_{\mu a}=\sum_i A_{\mu i}(\hat{x}_i-r^a_i),\hspace{0.5cm}
\tilde{d}_{\mu \alpha}=\sum_i A_{\mu i}(\hat{x}_i-|r^1_i|_0 x^{\alpha}_i).
\ee
Hereafter, we assume that the unspecified domain of integration denotes the integration over $\lsb -\infty:\infty \rsb$ or $\lsb -i \infty: i\infty \rsb$. We also assume that indices $a,b$ run over $1,\cdots,n$, and $\alpha,\beta$ over $1,\cdots,\nu$. The quantities $d_{\mu a}$ and $\tilde{d}_{\nu \alpha}$ consist of extensive sums of random variables $\lbb A_{\mu i} \rbb_{i}$, implying that the central limit theorem works and that $d$ and $\tilde{d}$ can be expressed by certain Gaussian variables. The mean is clearly zero, and the covariance becomes 
\be
\lsb d_{\mu a}d_{\omega b} \rsb_{A}
=
\delta_{\mu\omega}
\lb 
\frac{1}{N}\sum_{i}\hat{x}^2_{i}
-\frac{1}{N}\sum_{i} \hat{x}_{i}  r^{a}_{i}
-\frac{1}{N}\sum_{i} \hat{x}_{i} r^{b}_{i}
+\frac{1}{N}\sum_{i}r^{a}_{i}r^{b}_{i}
\rb.
\ee
Accordingly, we define the following order parameters:
\be
Q_{ab}=\frac{1}{N}\sum_{i}r^{a}_{i}r^{b}_{i},~m_{a}=\frac{1}{N}\sum_{i} \hat{x}_{i}  r^{a}_{i},
\ee
and assume the replica symmetry (RS) to be as follows:
\be
Q_{ab}=Q_1\delta_{ab}+q_1(1-\delta_{ab}),~ m_a=m_1.
\Leq{replica order parameters 1}
\ee
This RS assumption allows us to make many simplifications in dealing with $d_{\mu a}$, and $d_{\mu a}$ is represented by a sum of two independent Gaussian variables $v$ and $z$ drawn from $\mathcal{N}(0,1)$ as follows: 
\be
d_{\mu a}
=\sqrt{Q_1-q_1}v_{\mu a}+\sqrt{\hat{\rho} \sigma_x^2-2m_1+q_1}z_{\mu}
\equiv \sqrt{\Delta_1}v_{\mu a}+\sqrt{\Wh{M}_1}z_{\mu}.
\Leq{d}
\ee
Notice the relation $\hat{\rho}\sigma_x^2=(1/N)\sum_i \hat{x}_i^2$. A similar discussion and application of the RS are possible for the covariance of $\tilde{d}$:
\subbe
\Leq{replica order parameters 2}
\be
&&
\frac{1}{N}\sum_{i}|r^1_i|_0 x^{\alpha}_i x^{\beta}_i=Q_2\delta_{\alpha \beta}+q_2(1-\delta_{\alpha \beta}),
\\ &&
\frac{1}{N}\sum_{i}|r^1_i|_0 \hat{x}_i x^{\alpha}_i=m_2,
\\ &&
\frac{1}{N}\sum_{i} r_{i}^{a} |r^1_i|_0 x^{\alpha}_i =Q_c\delta_{a 1}+q_c(1-\delta_{a 1}). 
\ee
\subee
Using these covariances, we have a simple representation of $\tilde{d}$ as follows:
\be
\tilde{d}_{\mu \alpha}=\frac{\Delta_c}{\sqrt{\Delta_1}}v_{\mu 1} +\frac{\Wh{\mathcal{M}}_c}{\sqrt{\Wh{\mathcal{M}}_1}}z_{\mu }
+\sqrt{\Delta_2}u_{\mu \alpha}+\sqrt{\Wh{\mathcal{M}}_2-\frac{\Wh{\mathcal{M}}_c^2}{\Wh{\mathcal{M}}_1} -\frac{\Delta_c^2}{\Delta_1} }w_{\mu }.
\ee
where $u$ and $w$ denote new independent Gaussian variables from $\mathcal{N}(0,1)$, and $v$ and $z$ represent the same Gaussian variables as in \Req{d}; the following abbreviations are used:
\be
&&
\Wh{\mathcal{M}}_1=\hat{\rho} \sigma_x^2-2m_1+q_1,\hspace{0.2cm}
\\ &&
\Wh{\mathcal{M}}_2=\hat{\rho} \sigma_x^2-2m_2+q_2,\hspace{0.2cm}
\\ &&
\Wh{\mathcal{M}}_c=\hat{\rho} \sigma_x^2-(m_1+m_2)+q_c,
\\ &&
\Delta_1=Q_1-q_1,\hspace{1cm}
\Delta_2=Q_2-q_2,\hspace{1cm}
\Delta_c=Q_c-q_c.
\ee
The above order parameters are nothing but those having the same symbols in \Req{order parameters}, but now, they are represented using replicas. Note that in the limit $\beta \to \infty$, all $\Delta_{1,2,c}$ vanish and $\Wh{\mathcal{M}}_{1,2,c}$ converge to $\mathcal{M}_{1,2,c}$. 

On the basis of the above consideration and denoting the weight of normal distribution $\mathcal{N}(0,1)$ as $Dx=e^{-\frac{1}{2} x^2}/\sqrt{2\pi}$, we obtain the following expression:
\be
\lsb  
 e^{
 -\frac{1}{2}\beta  
 \lbb 
 \sum_{a=1}^{n}  \lb d_{\mu a} +\xi_{\mu} \rb^2 + \sum_{\alpha=1}^{m}   \lb \tilde{d}_{\mu \alpha}+\xi_{\mu} \rb^2
\rbb
}
\rsb_{\V{\xi},A} \equiv L^M,
\ee
where
\be
L=
\int D\eta  Dz Dw  \lb \int Dv~e^{-\frac{1}{2} \beta h^2_1(v,z,\eta) } \rb ^n
\Ave{ \lb \int Du~e^{-\frac{1}{2}\beta h^2_2(v,z,u,w,\eta)} \rb^{\nu}  }_{v|h_1},
\ee
where 
\be
&&
h_1(v,z,\eta)=\sqrt{\Delta_1}v+\sqrt{\Wh{\mathcal{M}}_1}z+\sigma_{\xi}^2 \eta,
\\ &&
h_2(v,z,u,w,\eta)=\frac{\Delta_c}{\sqrt{\Delta_1}}v +\frac{\Wh{\mathcal{M}}_c}{\sqrt{\Wh{\mathcal{M}}_1}}z
+\sqrt{\Delta_2}u+\sqrt{\Wh{\mathcal{M}}_2-\frac{\Wh{\mathcal{M}}_c^2}{\Wh{\mathcal{M}}_1} -\frac{\Delta_c^2}{\Delta_1} }w +\sigma_{\xi}^2 \eta,
\\ &&
\Ave{ \cdots }_{v|h_1}=\frac{\int Dv (\cdots )e^{-\frac{1}{2} \beta h^2_1(v,z,\eta) }}{\int Dv~e^{-\frac{1}{2} \beta h^2_1(v,z,\eta) }}.
\ee
To proceed with the calculations, we use a trick to perform a trace over $\V{r}$ and $\V{x}$: rewriting the order parameters as integration variables and introducing delta functions that require order parameters to take the values defined in \Reqs{replica order parameters 1}{replica order parameters 2}. This yields the following:
\be
\Phi_{2}=\int d\Omega~I L^{M},
\ee
where $\Omega=\lbb Q_1,q_1,m_1,Q_2,q_2,m_2,Q_c,q_c,\rbb$ and 
\be
&&
I=
 \Wt{\Tr{\lbb \V{r}_a \rbb_{a=1}^{n} }}
  \Tr{\lbb \V{x}_{\alpha} | \V{r}_1 \rbb_{\alpha=1}^{\nu} }
\prod_{a=1}^{n} \delta(NQ_1-\sum_i (r^a_i)^2 )
\prod_{a<b} \delta(Nq_1-\sum_i r^a_i r^b_i )
\prod_{a=1}^{n} \delta(Nm_1-\sum_i  \hat{x}_i r^a_i )
\no \\ &&
\prod_{\alpha=1}^{n} \delta(NQ_2-\sum_i  | r^1_i |_0 (x^{\alpha}_i)^2 )
\prod_{\alpha<\beta}  \delta(Nq_2-\sum_i  | r^1_i |_0 x^{\alpha}_ix^{\beta}_i )
\prod_{a=1}^{n} \delta(Nm_2-\sum_i  \hat{x}_i | r^1_i |_0 x^{\alpha}_i )
\no \\ &&
\prod_{\alpha=1}^{\nu}\delta(NQ_c-\sum_i  r^1_i  x^{\alpha}_i ) 
\prod_{a=2}^{n}\prod_{\alpha=1}^{\nu}\delta(Nq_c-\sum_i  | r^1_i |_0 r^a_i x^{\alpha}_i ).
\ee
We rewrite these delta functions by using the Fourier representations. In doing so, constant factors can be applied to the Fourier integration variables, and we choose convenient factors for later calculations. For example, the delta functions of $Q_1$ and $q_1$ are expressed as follows:
\be
&&
\delta \lb NQ_1-\sum_{i}\lb r_i^{a}\rb^2 \rb
=C_1 \int d\hat{Q}_1~e^{
\frac{1}{2}NQ_1\hat{Q}_1-\frac{1}{2}\hat{Q}_1\sum_{i}( x_i^{a \alpha } )^2
}
\\ &&
\delta \lb Nq_1-\sum_{i}r_i^{a}r_i^{b} \rb
=c_1 \int d\hat{q}_1~e^{
-Nq_1\hat{q}_1+\hat{q}_1\sum_{i}r_i^{a }r_i^{b}
},
\ee
where $C_1$ and $c_1$ denote the normalization factors but will be discarded hereafter because they do not contribute in the limit $N\to \infty$. These operations provide the following:
  \be
I=\int d \hat{\Omega}~
e^{N
\lb 
S+
\lsb \log 
\Wt{ \Tr{ \{r_a\}_a  } }  \Tr{ \{x_{\alpha}| r_1 \}_{\alpha} }   e^{f(r,x|\hat{x} )} 
\rsb_{ \hat{x} }
 \rb
 },
\ee
where $\hat{\Omega}=\lbb \hat{Q}_1,\hat{q}_1,\hat{m}_1,\hat{Q}_2,\hat{q}_2,\hat{m}_2,\hat{Q}_c,\hat{q}_c,\rbb$ and
\be
&&
S=
\frac{1}{2}n Q_1\hat{Q}_1 - \frac{1}{2}n(n-1) q_1\hat{q}_1
+
\frac{1}{2}\nu Q_2\hat{Q}_2 - \frac{1}{2}\nu(\nu-1) q_2\hat{q}_2 
\no \\ &&
-n m_1 \hat{m}_1-\nu m_2 \hat{m}_2
-\nu Q_c \hat{Q}_c-\nu (n-1)q_c \hat{q}_c,
\\
&&
\hspace{-2cm}
f(r,x|\hat{x})
= 
-\frac{1}{2}\hat{Q}_1\sum_{a}r_a^2+\hat{q}_1 \sum_{a<b}r_a r_b +\hat{m}_1 \hat{x}\sum_{a}r_a
\no \\ &&
\hspace{-2cm}
+|r_{1}|_0\lbb 
-\frac{1}{2}\hat{Q}_2\sum_{\alpha}x_{\alpha}^2+\hat{q}_1 \sum_{\alpha<\beta }x_{\alpha}x_{\beta} 
+\hat{m}_2 \hat{x}\sum_{\alpha}x_{\alpha}
+\hat{\Delta}_cr_1 \sum_{\alpha}x_{\alpha}+\hat{q}_c \lb \sum_{a=1}^{n}r_a \rb \lb \sum_{\alpha} x_\alpha \rb
\rbb,
\ee
where we set $\hat{\Delta}_c=\hat{Q}_c-\hat{q}_c$. The average over $\hat{x}$ appears as a result of the law of large numbers. As noted in the main text, we consider the Bernoulli--Gaussian distribution with respect to $\hat{x}$. Denoting its Gaussian part as $P_{G}(\hat{x})=\exp(-\frac{\hat{x}^2}{2\sigma_x^2} )/\sqrt{2\pi \sigma_x^2}$, we obtain the following:
\be
\lsb \log 
\Wt{ \Tr{ \{r_a\}_a  } }  \Tr{ \{x_{\alpha}| r_1 \}_{\alpha} }   e^{f(r,x|\hat{c},\hat{x} )} 
\rsb_{ \hat{x} }
=\hat{\rho}\int d\hat{x}~P_{G}(\hat{x})\log J_{A}+(1-\hat{\rho})\log J_{I},
\ee
where 
\be
J_{A}
\equiv
\Wt{ \Tr{ \{r_a\}_a  } }  \Tr{ \{x_{\alpha}| r_1 \}_{\alpha} }   e^{f(r,x|\hat{x} )},~~
J_{I}
\equiv
\Wt{ \Tr{ \{r_a\}_a  } }  \Tr{ \{x_{\alpha}| r_1 \}_{\alpha} }   e^{f(r,x|0)}.
\ee
Cross-quadratic terms in $f$ can be transformed into linear terms. First, they are transformed as follows: 
\be
\sum_{a<b}r_a r_b =\frac{1}{2} 
\lbb 
  \lb \sum_{a}r_a \rb^2 - \sum_{a}r_a^2
\rbb
,~
\sum_{\alpha<\beta} |r_1|_0 x_{\alpha} x_{\beta}
=\frac{1}{2}
\lbb 
\lb \sum_{\alpha} | r_1 |_0 x_{ \alpha } \rb^2 - \sum_{\alpha} \lb  |r_1|_0 x_{\alpha}  \rb^2 
\rbb.
\ee
Note that $| \cdot  |_0^k=| \cdot |_0 ~(k > 0)$. Let us set $X\equiv \sum_{a}r_a$ and $Y\equiv \sum_{\alpha}|r_{1}|_0 x_{\alpha}$. The minimum number of auxiliary variables to break the quadratic terms is two, but here, we introduce three auxiliary variables to make the resultant formula interpretable. Accordingly, we have the following: 
\be
&&
\int Dv Du Dw~e^{(v,u,w) \cdot (aX,bY, cX+dY)^{t} }=e^{\frac{1}{2} \lb (a^2+c^2)X^2+(b^2+d^2)Y^2 +2cd XY   \rb}
\no \\ &&
=
e^{
\frac{1}{2} \lb 
q_1 (\sum_{a}r_a)^2 +
 \hat{q}_1 (\sum_{\alpha} |r_{1}|_0 x_{\alpha})^2 
+2\hat{q}_c \lb \sum_{a=1}^{n}r_a \rb  \lb \sum_{\alpha} |r_{1}|_0 x_\alpha  \rb 
\rb
}.
\ee
A simple solution of this equation with respect to $a,b,c$, and $d$ is as follows:
\be
a=\sqrt{ \hat{q}_1-\hat{q}_c },~
b=\sqrt{ \hat{q}_2-\hat{q}_c },~
c=d=\sqrt{ \hat{q}_c }.
\ee
Hence, we can set $J_{A}=\Wt{ \Tr{ \{r_a\}_a  } }  \Tr{ \{x_{\alpha}| r_1 \}_{\alpha} } \int Dv Du Dw~e^{g_{A}}$ with 
\be
&&
\hspace{-1.8cm}
g_{A} =
-\frac{1}{2}\lb \hat{Q}_1+\hat{q}_1 \rb\sum_{a}r_a^2 +A_1\sum_{a}r_a 
+
|r_{1}|_0 \lb 
-\frac{1}{2}\lb \hat{Q}_2+\hat{q}_2 \rb\sum_{\alpha}x_{\alpha}^2 +(A_2+\hat{\Delta}_c r_1)\sum_{\alpha}x_{\alpha} 
\rb, 
\\
&&
\hspace{0cm}
A_1(\hat{m}_1)=\sqrt{ \hat{q}_1-\hat{q}_c } v+\sqrt{ \hat{q}_c }w+\hat{m}_1\hat{x},~
A_2(\hat{m}_2)=\sqrt{ \hat{q}_2-\hat{q}_c } u+\sqrt{ \hat{q}_c }w+\hat{m}_2\hat{x}.
\ee
This formula is nice because $g$ is expected to be $O(\beta)$ in the $\beta \to \infty$ limit: $\hat{q}$ and $\hat{Q}$ are $O(\beta^2)$; $(\hat{Q}+\hat{q})$ and $\hat{\Delta}_c=(\hat{Q}_c-\hat{q}_c)$  are $O(\beta)$. Now, we can easily perform the integration over $x$ as follows:
\be
 \Tr{ \{x_{\alpha}| r_1 \}_{\alpha} }   e^{g_{A } } 
 =
 e^{
 -\frac{1}{2}\lb \hat{Q}_1+q_1 \rb\sum_{a}r_a^2 +A_1(\hat{m}_1)\sum_{a}r_a 
 +\frac{\nu}{2}|r_1|_0 \lb \frac{\lb A_2(\hat{m}_2)+\hat{\Delta}_c r_1 \rb^2}{\hat{Q}_2+\hat{q}_2}+\log \frac{2\pi}{\hat{Q}_2+\hat{q}_2}\rb
 },
\ee 
and thus, 
\be
J_{A}=
\int Dv Du Dw~\lb \Wt{\Tr{r}} e^{ -\frac{1}{2}(\hat{Q}_1+\hat{q}_1) r^2+A_1(\hat{m}_1) r} \rb^n
\Ave{
 e^{
 \frac{\nu}{2}|r_1|_0 \lb \frac{\lb A_2(\hat{m}_2)+\hat{\Delta}_c r_1 \rb^2}{\hat{Q}_2+\hat{q}_2}+\log \frac{2\pi}{\hat{Q}_2+\hat{q}_2}\rb
 }
}_{r_1| \hat{m}_1 },
\ee
where 
\be
\Ave{\cdots }_{r_1| \hat{m}_1 } =\frac{
\Wt{\Tr{r_1}}
(\cdots)
e^{ -\frac{1}{2}(\hat{Q}_1+\hat{q}_1) r^2_1+A_1(\hat{m}_1) r_1} 
}{
\Wt{\Tr{r_1}} e^{ -\frac{1}{2}(\hat{Q}_1+\hat{q}_1) r^2_1+A_1(\hat{m}_1) r_1} 
}.
\ee
Setting $\hat{m}_1=\hat{m}_2=0$ in $J_{A}$, we have the expression of $J_I$. Hence,
\be
&& \hspace{-1cm}
\phi_2(n,\nu,\beta )\equiv \frac{1}{N}\log \Phi(n,\nu,\beta )=\alpha \log L+\frac{1}{N}\log I
\no \\ && \hspace{-1cm}
=
\alpha \log \int D\eta Dz Dw  \lb \int Dv~e^{-\frac{1}{2} \beta h^2_1(v,z,\eta) } \rb ^n
\Ave{ \lb \int Du~e^{-\frac{1}{2}\beta h^2_2(v,z,u,w,\eta)} \rb^\nu   }_{v|h_1}
\no \\ && \hspace{-1cm}
+\frac{1}{2}n Q_1\hat{Q}_1 - \frac{1}{2}n(n-1) q_1\hat{q}_1
+\frac{1}{2}\nu  Q_2\hat{Q}_2 - \frac{1}{2}\nu (\nu -1) q_2\hat{q}_2 
-nm_1 \hat{m}_1-\nu m_2 \hat{m}_2
-\nu Q_c \hat{Q}_c-\nu (n-1)q_c \hat{q}_c
\no \\ && \hspace{-1cm}
+
\hat{\rho}\int d \hat{x}~P_{G}(\hat{x})
\log  \lbb
\int Dv Du Dw~\lb \Wt{\Tr{r}} e^{ -\frac{1}{2}(\hat{Q}_1+\hat{q}_1) r^2+A_1(\hat{m}_1) r} \rb^n
\Ave{
 e^{
 \frac{\nu }{2}|r_1|_0 \lb \frac{\lb A_2(\hat{m}_2)+\hat{\Delta}_c r_1 \rb^2}{\hat{Q}_2+\hat{q}_2}+\log \frac{2\pi}{\hat{Q}_2+\hat{q}_2}\rb
 }
}_{r_1| \hat{m}_1 }
\rbb
\no \\ && \hspace{-1cm}
+(1-\hat{\rho})
\log \lbb 
\int Dv Du Dw~\lb \Wt{\Tr{r}} e^{ -\frac{1}{2}(\hat{Q}_1+\hat{q}_1) r^2+A_1(0) r} \rb^n
\Ave{
 e^{
 \frac{\nu }{2}|r_1|_0 \lb \frac{\lb A_2(0)+\hat{\Delta}_c r_1 \rb^2}{\hat{Q}_2+\hat{q}_2}+\log \frac{2\pi}{\hat{Q}_2+\hat{q}_2}\rb
 }
}_{r_1| 0}
\rbb.
\ee
Let us glance at the interdependency of the order parameters. Let us set $\tilde{\Omega}_1=\lbb Q_1,q_1,m_1,\hat{Q}_1,\hat{q}_1,\hat{m}_1 \rbb$, $\tilde{\Omega}_2=\lbb Q_2,q_2,m_2,\hat{Q}_2,\hat{q}_2,\hat{m}_2 \rbb$, and $\tilde{\Omega}_c=\lbb Q_c,q_c,\hat{Q}_c,\hat{q}_c,\rbb$. We see that
\be
\phi_2(n,0,\beta)=F_1(\tilde{\Omega}_1),~ 
\phi_2(0,\nu,\beta)=F_2(\tilde{\Omega}_1,\tilde{\Omega}_2,\tilde{\Omega}_c),~
\phi_2(0,0,\beta)=0.
\ee 
This equation implies that $\phi_2(n,\nu)$ has multiple solutions in the saddle-point equation of the order parameters at and around $n=\nu=0$. We should choose a solution that is analytically continued to the solution at $\nu=0$  with respect to $\nu$. Hence, we first take the $\nu \to 0$ limit, yielding $\phi_1(n,\beta) \equiv (1/N)\log \Phi_1(n,\beta)=\phi_2(n,0,\beta)$. 

\subsection{Derivation of $f_1$}\Lsec{Derivation of f_1}
The free energy $f_1$ is obtained from $\phi_1$ to $-\beta f_1=\lim_{n\to 0}(1/n)\phi_1(n,\beta)$. Performing the variable transformation $(\sqrt{\hat{q}_1}v+\hat{m}_1\hat{x})/(\sqrt{\hat{q}_+\hat{m}_1^2\sigma_x^2})\to z$, we obtain the following:
\be
&&
-\beta f_1 =\Extr{\tilde{\Omega}_1}
\Biggl\{
\frac{1}{2} \hat{Q}_{1}Q_{1}+\frac{1}{2} \hat{q}_{1}q_{1}- \hat{m}_{1}m_{1}
+\hat{\rho} \int Dz \log X_{A} + ( 1-\hat{\rho} ) \int Dz \log  X_{I}
\no \\ &&
-\frac{\alpha}{2} 
\lb
\log \lb 1+\beta\Delta_1 \rb
+\frac{\beta (\Wh{M}_1 +\sigma_{\xi}^2)}{1+\beta\Delta_1}
\rb
\Biggr\},
\ee
where  
\be
&&
X_{A}=
\int dx~e^{-\frac{1}{2}(\hat{Q}_{1}+\hat{q}_{1})x^2+\sqrt{ \hat{q}_{1}+ \hat{m}_{1}^2\sigma_{x}^2 }~z x -\beta \lambda |x|  }, 
\ee
and the expression of $X_{I}$ is obtained by setting $\hat{m}_1=0$ in $X_{A}$. To take the zero-temperature limit $\beta \to \infty$, we assume the following scalings:
\subbe
\be
&&
\beta \Delta_1\to \chi_1,
\\ &&
(\hat{Q}_1+\hat{q}_1)\to \beta \hat{Q}_1,
\\ &&
\hat{q}_1 \to \beta^2 \hat{\chi}_1 ,
\\ &&
\hat{Q}_1\to -\beta^2 \hat{\chi}_1,
\\ &&
 \hat{m}_1\to \beta \hat{m}_1.
\ee
\subee
Then, the integration is dominated by the saddle point $x^*$ in the limit $\beta \to \infty$
\be
X_{A}
\to 
e^{\beta f_{A}(x^*)},
\ee
where
\be
f_{A }(x)=-\frac{1}{2}\hat{Q}_1x^2 +
\left\{
\begin{array}{cc}
A_+ x  & (x \ge 0)    \\
A_- x &   (x<0)  
\end{array}
\right.,
\ee
and
\be
A_{\pm}=\sqrt{\hat{\chi}_1 +\hat{m}_1^2\sigma_x^2 } z \mp \lambda.
\ee
The saddle point $x^*$ changes the behavior depending on the value of $A_{\pm}$. Simple algebra yields the following:
\be
\lim_{\beta \to \infty} \frac{1}{\beta}\int Dz \log X_A=
\int Dz f_{\hat{p}}(x^*)=\frac{F(\theta_A) }{\hat{Q}_{1}}.
\ee 
A similar calculation is possible for $X_I$. Summarizing the calculations, we obtain \Req{f_1-zerotemp}.

\subsection{Derivation of $f_2$}\Lsec{Derivation of f_2}
A small calculation from $\phi_2$ yields the following: 
\be
&& \hspace{-1.cm}
-\beta f_2=\lim_{\nu \to 0} \frac{1}{\nu}\phi_2(0,\nu,\beta)
\no \\ && \hspace{-1.cm}
=
\alpha \int D\eta Dz Dw  
\Ave{ \log \int Du~e^{-\frac{1}{2}\beta h^2_2(v,z,u,w,\eta)}   }_{v|h_1}
+\frac{1}{2} Q_2\hat{Q}_2 + \frac{1}{2} q_2\hat{q}_2 
-m_2 \hat{m}_2
-Q_c \hat{Q}_c+q_c \hat{q}_c
\no \\ && \hspace{-1cm}
+\frac{\hat{\rho}}{2}
\int d \hat{x}~P_{G}(\hat{x})\int Dv Du Dw~
\Ave{
|r_1|_0 \lb \frac{\lb A_2(\hat{m}_2)+\hat{\Delta}_c r_1 \rb^2}{\hat{Q}_2+\hat{q}_2}+\log \frac{2\pi}{\hat{Q}_2+\hat{q}_2}\rb
}_{r_1| \hat{m}_1 }
\no \\ && \hspace{-1cm}
+\frac{1-\hat{\rho}}{2}
\int Dv Du Dw~
\Ave{
|r_1|_0 
\lb \frac{\lb A_2(0)+\hat{\Delta}_c r_1 \rb^2}{\hat{Q}_2+\hat{q}_2}+\log \frac{2\pi}{\hat{Q}_2+\hat{q}_2}\rb
}_{r_1| 0}.
\ee
We assume the following scalings:
\subbe
\be
&&
\beta (Q_1-q_1)\to \chi_1,~ \beta (Q_2-q_2)\to \chi_2,~ \beta (Q_c-q_c)\to \chi_c,~
\\ &&
(\hat{Q}_1+\hat{q}_1)\to \beta \hat{Q}_1,~ (\hat{Q}_2+\hat{q}_2)\to \beta \hat{Q}_2,~ (\hat{Q}_c-\hat{q}_c)\to \beta \hat{Q}_c,
\\ &&
\hat{q}_1,-\hat{Q}_1 \to \beta^2 \hat{\chi}_1,~\hat{q}_2,-\hat{Q}_2 \to \beta^2 \hat{\chi}_2,~\hat{q}_c,\hat{Q}_c \to \beta^2 \hat{\chi}_c,
\\ &&
\hat{m}_1\to \beta \hat{m}_1,~\hat{m}_2\to \beta \hat{m}_2.
\ee
\subee
After lengthy but straightforward calculations, we obtain the following:
\be
&&
\lim_{\beta \to \infty} 
\frac{1}{\beta} \int D\eta Dz Dw  
\Ave{ \log \int Du~e^{-\frac{1}{2}\beta h^2_2(v,z,u,w,\eta )}   }_{v|h_1}
\no \\ &&
=-\frac{1}{2}\frac{1}{1+\chi_2}
\lbb 
\frac{\chi_c^2}{(1+\chi_1)^2}\Wt{\mathcal{M}}_1  -2\frac{\chi_c}{1+\chi_1}\Wt{\mathcal{M}}_c+\Wt{\mathcal{M}}_2  
\rbb, 
\ee
and
\be
&&
\lim_{\beta \to \infty} \frac{1}{\beta}
\int d \hat{x}~P_{G}(\hat{x})\int Dv Du Dw~
\Ave{
|r_1|_0 \lb \frac{\lb A_2(\hat{m}_2)+\hat{\Delta}_c r_1 \rb^2}{\hat{Q}_2+\hat{q}_2}+\log \frac{2\pi}{\hat{Q}_2+\hat{q}_2}\rb
}_{r_1| \hat{m}_1 }
\no \\ &&
=\frac{1}{\hat{Q}_2} 
\int d \hat{x}~P_{G}(\hat{x})\int Dv Du Dw~
|r^*_1|_0 \lb \sqrt{\hat{\chi}_2-\hat{\chi}_c}u +\sqrt{\hat{\chi}_c}w+\hat{m}_2\hat{x}+\hat{Q}_cr^*_1  \rb^2.
\ee
The saddle point $r^*_1$ depends on $v,w$, and $\hat{x}$, and we need to be careful while evaluating it. Let us set and expand
\be
&&
T=
\int d \hat{x}~P_{G}(\hat{x})\int Dv Du Dw~
|r^*_1|_0 \lb \sqrt{\hat{\chi}_2-\hat{\chi}_c}u +\sqrt{\hat{\chi}_c}w+\hat{m}_2\hat{x}+\hat{Q}_cr^*_1  \rb^2
\no \\ &&
=(\hat{\chi}_2-\hat{\chi}_c) \int d \hat{x}~P_{G}(\hat{x})\int Dv Dw~ |r_1^*(v,w,\hat{x})|_0+
\int d \hat{x}~P_{G}(\hat{x})\int Dv  Dw~R,
\ee
where the integration of $u$ is easily performed since it is independent of $r_1^*$, and 
\be
R=|r^*_1(v,w,\hat{x})|_0\lb \sqrt{\hat{\chi}_c}z+\hat{m}_2\hat{x}+\hat{Q}_cr^*_1  \rb^2.
\ee
The first integral is evaluated as follows:
\be
\int d \hat{x}~P_{G}(\hat{x})\int Dv Dw~|r_1(v,w,\hat{x})|_0=2E_0(\theta_{A}).
\ee
This can be shown by changing the integration variable as $\sqrt{\hat{\chi}_1-\hat{\chi}_c}v+\sqrt{\hat{\chi}_c}w+\hat{m}_1\hat{x} \to \sqrt{\hat{\chi}_1+\hat{m}_1^2\sigma_x^2 }z$, which appears in $\Ave{\cdots}_{r_1|\hat{m}_1}$. Each term of $R$ is calculated in a similar manner by performing Gaussian integrations many times. Here, we summarize the result:
\be
&&
X_1=
\int d \hat{x}~P_{G}(\hat{x})\int Dv Dw~|r_1^*|_0~w^2
=
2E_0(\theta_{A} )+2\frac{\hat{\chi}_c}{\hat{\chi}_1+\hat{m}_1^2\sigma_x^2}
\frac{ \theta_{A} }{ \sqrt{2\pi} }e^{-\frac{1}{2}\theta_{A}^2 },
\\ &&
X_2 = \int d \hat{x}~P_{G}(\hat{x})\int Dv Dw~|r_1^*|_0~w\hat{x}.
=2\frac{\hat{m}_1\sigma_x^2 \sqrt{\chi_c}}{\hat{\chi}_1+\hat{m}_1^2\sigma_x^2 } 
\frac{ \theta_{A} }{\sqrt{2\pi} }e^{-\frac{1}{2}\theta_{ A }^2 },
\\ &&
X_3=\int d \hat{x}~P_{G}(\hat{x})\int Dv Dw~|r_1^*|_0~\hat{x}^2
=
2 \sigma_x^2 E_0(\theta_{A} )
+
2\frac{ \hat{m}_1^2  \sigma_{x}^4 }{ \hat{\chi}_1+\hat{m}_1^2\sigma_x^2 }
\frac{ \theta_{A} }{ \sqrt{2\pi} }e^{-\frac{1}{2}\theta_{A}^2 },
\\ &&
X_4=\int d \hat{x}~P_{G}(\hat{x})\int Dv Dw~w r_1^*
=
2\frac{\sqrt{\chi_c} }{\hat{Q}_1}E_0(\theta_{A} ),
\\ &&
X_5
=\int d \hat{x}~P_{G}(\hat{x})\int Dv Dw~ \hat{x} r_1^* 
=2\frac{\hat{m}_1\sigma_x^2 }{\hat{Q}_1}E_0(\theta_{A} ),
\\ &&
X_6
=\int d \hat{x}~P_{G}(\hat{x})\int Dv Dw~ (r_1^*)^2 
=2\frac{1}{ \hat{Q}_1^2 }F(\theta_{A}),
\ee
Collecting all the terms and rewriting them with \Req{EOS-ell_1}, we finally obtain \Req{epsilon_2}.



\begin{thebibliography}{99}

\bibitem{Okada:13} http:\slash\slash{}sparse-modeling.jp\slash{}index$\_$e.html

\bibitem{Rish:14} I. Rish and G. Grabarnik, {\it Sparse Modeling: Theory, Algorithms, and Applications}, (CRC Press, 2014)

\bibitem{Mairal:14} J. Mairal, F. Bach, and J. Ponce, Sparse Modeling for Image and Vision Processing, arXiv:1411.3230v2 

\bibitem{Hastie:15} T. Hastie, R. Tibshirani, and M. Wainwright, {\it Statistical Learning with Sparsity: The Lasso and Generalizations}, (CRC Press, 2015)

\bibitem{Tibshirani:96} R. Tibshirani, Regression shrinkage and selection via the lasso. {\it J. Royal. Statist. Soc B.}, {\bf 58}, 267--288  (1996)

\bibitem{Efron:04} B. Efron, T. Hastie, I. Johnstone, and R. Tibshirani, LEAST ANGLE REGRESSION, {\it The Annals of Statistics}, {\bf 32}, 407--499, (2004)

\bibitem{Wright:09} J. Wright, A. Y. Yang, A. Ganesh,  S. . Sastry, and Y. Ma, Robust Face Recognition via Sparse Representation, {\it IEEE TRANSACTIONS ON PATTERN ANALYSIS AND MACHINE INTELLIGENCE}, {\bf 31}, (2009)

\bibitem{Elith:06} J. Elith, C. H. Graham, R. P. Anderson, et al. Novel methods improve prediction of species' distributions from occurrence data, {\it ECOGRAPHY} {\bf 29}, 129--151, (2006)

\bibitem{Schafer:05} J. Sch\"afer and K. Strimmer, A shrinkage approach to large-scale covariance matrix estimation and implications for functional genomics, {\it STATISTICAL APPLICATIONS IN GENETICS AND MOLECULAR BIOLOGY}, {\bf 4}, (2005)

\bibitem{Kato:12} T. Kato and M. Uemura, Period Analysis using the Least Absolute Shrinkage and Selection Operator (Lasso), {\it Publ. Astron. Soc. Japan}, {\bf 64}, (2012)

\bibitem{Uemura:15} M. Uemura, K. S. Kawabata, S. Ikeda, K. Maeda: Variable selection for modeling the absolute magnitude at maximum of Type Ia supernovae, {\it Publ. Astron. Soc. Japan}, {\bf 67}, 55, 1--9, (2015).


\bibitem{Donoho:06} D. L. Donoho, {\it IEEE Transactions on Information Theory}, {\bf 52}(4), 1289--1306, (2006).

\bibitem{Candes:05} E. J. Cand\`es and T. Tao, {\it IEEE Transactions on Information Theory}, {\bf 51}(12), 4203--4215, (2005).

\bibitem{Candes:06a} E. J. Cand\`es, J. Romberg and T. Tao, {\it IEEE Transactions on Information Theory}, {\bf 52}(2), 489--509, (2006).

\bibitem{Candes:06b} E. J. Cand\`es and T. Tao, {\it IEEE Transactions on Information Theory}, {\bf 52}(12), 5406--5425, (2006).

\bibitem{Donoho:09-1} D. L.  Donoho and J. Tanner: {\it Phil. Trans. R. Soc. A}, {\bf 367}, 4273--4293, (2009)

\bibitem{Donoho:09-2} D. L. Donoho, A. Malekib, and A. Montanari: Message-passing algorithms for compressed sensing, {\it Proc. Natl. Acad. Sci.}, {\bf 106}, 18914--18919, (2009)

\bibitem{Kabashima:09} Y. Kabashima, T. Wadayama, and T. Tanaka: A typical reconstruction limit for compressed sensing based on $L_p$-norm minimization, {\it J. Stat. Mech.}, L09003, (2009)

\bibitem{Ganguli:10} S. Ganguli and H. Sompolinsky: Statistical Mechanics of Compressed Sensing, {\it Phys. Rev. Lett.}, {\bf 104}, 188701, (2010)

\bibitem{Rangan:10} S. Rangan, Generalized Approximate Message Passing for Estimation with Random Linear Mixing,
 {\it arXiv:1010.5141}, (2010) 
 
 \bibitem{Krzakala:12} F. Krzakala, M. M\'ezard, F. Sausset, Y. Sun and L. Zdeborov\'a, Probabilistic reconstruction in compressed sensing: algorithms, phase diagrams, and threshold achieving matrices, {\it J. Stat. Mech.}, P08009,  (2012)

\bibitem{Sakata:13} A. Sakata and Y. Kabashima, {\it Europhysics Letters}, {\bf 103}, 28008--p1--28008--p6, (2013)

\bibitem{Nakanishi:15} Y. Nakanishi, T. Obuchi, M. Okada, and Y. Kabashima, arXiv:1510.02189, to appear in {\it J. Stat. Mech.}

\bibitem{Yedidia:03} J. S. Yedidia, W. T. Freeman, and Y Weiss, {\it Understanding belief propagation and its generalizations} (in Exploring artificial intelligence in the new millennium, pp. 239--269, Morgan Kaufmann Publishers Inc., San Francisco, CA, 2003)

\bibitem{ADVA}  M. Opper and D. Saad, {\it Advanced Mean Field Methods: Theory and Practice}, (Neural Information Processing series, A Bradford Book, 2001)

\bibitem{Seung:92} H. S. Seung, H. Sompolinsky, and N. Tishby: Statistical mechanics of learning from examples, {\it Phys. Rev. A}, {\bf 45}, 6056-6091, (1992)

\bibitem{Opper:96} M. Opper and O. Winther, A mean field algorithm for Bayes learning in large feed-forward neural networks, {\it Advances in Neural Information Processing Systems 9}, NIPS, (1996)

\bibitem{Kabashima:03} Y. Kabashima, A CDMA multiuser detection algorithm on the basis of belief propagation, {\it J. Phys. A}, {\bf 36}, 11111--11121, (2003) 

\bibitem{Bayati:10} M. Bayati, J. Bento, and A. Montanari, The LASSO risk: asymptotic results and real world
examples, {\it Advances in Neural Information Processing Systems 23}, NIPS, (2010)

\bibitem{Homrighausen:14} D. Homrighausen and D. J. McDonald, Leave-one-out cross-validation is risk consistent for lasso, {\it Mach. Learn.} {\bf 97}, 65--78, (2014)

\bibitem{Berkeley} http:\slash\slash{}hercules.berkeley.edu\slash{}database\slash{}index$\_$public.html

\bibitem{Silverman:12} J. M. Silverman, M. Ganeshalingam, W. Li, and A. V. Filippenko, Berkeley Supernova Ia Program -- III. Spectra near maximum brightness improve the accuracy of derived distances to Type La supernovae, {\it Mon. Not. R. Astron. Soc.}, {\bf 425}, 1889--1916, (2012)

\bibitem{Xu:12} H. Xu and S. Mannor, Sparse Algorithms Are Not Stable: A No-Free-Lunch Theorem, {\it IEEE Transactions on pattern analysis and machine intelligence}, {\bf 34}, 187--193, (2012)
\bibitem{The Elements} T. Hastie, R. Tibshirani, and J. Friedman, {\it The Elements of Statistical Learning: Data Mining, Inference, and Prediction}, (Springer Series in Statistics, 2009)

\bibitem{Classification}  L. Breiman, J. Friedman, C. J. Stone, and R.A. Olshen, {\it Classification and Regression Trees}, (Chapman and Hall/CRC, 1984) 


\bibitem{Tibshirani:01} R. Tibshirani, G. Walther and T. Hastie, Estimating the Number of Clusters in a Data Set via the Gap Statistic, {\it Journal of the Royal Statistical Society}, {\bf 63}, 411--423, (2001)  



\end{thebibliography}
\end{document}